\documentclass[
aps,
prb,
showpacs,
floatfix,
reprint,
twoside,
superscriptaddress
]{revtex4-1}
\usepackage{amsmath}
\usepackage{amssymb}
\usepackage[T1]{fontenc}
\usepackage{fourier}
\usepackage{graphicx}
\usepackage[colorlinks=true,
allcolors=blue,
dvipdfm=true]{hyperref}

\begin{document}

\title{Static and dynamic properties of Josephson weak links with singlet and triplet coupling}

\author{Andreas Moor}
\affiliation{Theoretische Physik III, Ruhr-Universit\"{a}t Bochum, D-44780 Bochum, Germany}
\author{Anatoly F.~Volkov}
\affiliation{Theoretische Physik III, Ruhr-Universit\"{a}t Bochum, D-44780 Bochum, Germany}

\begin{abstract}
We theoretically study static and dynamic properties of short Josephson junctions (JJ) with singlet and triplet Josephson coupling. In singlet Josephson weak links, two singlet superconductors~S are connected with each other by a normal film~(N) or wire. Triplet JJs, which we denote~S$_{\text{m}}$-N(F)-S$_{\text{m}}$, are formed by two singlet BCS superconductors covered by a thin layer of a weak ferromagnet~F$_{\text{w}}$. These superconductors~S$_{\text{m}}$ are separated from the~N (or~F) layer by spin filters, which pass electrons with only one spin orientation. The triplet Cooper pairs propagating from the left (right) superconductors~S$_{\text{m}}$ differ from each other not only by polarizations, but also by chiralities. The latter is determined by the magnetization orientation in weak ferromagnets~F$_{\text{w}}$. We obtain analytical formulas for the critical Josephson current in both types of the JJs. If chiralities of the triplet Cooper pairs penetrating into the~N film in S$_{\text{m}}$-N(F)-S$_{\text{m}}$ JJs from the left and right~S$_{\text{m}}$ are different, the Josephson current is not zero in the absence of the phase difference (spontaneous Josephson current). We also calculate the admittance~$Y(\Omega)$ for arbitrary frequencies~$\Omega$ in the case of singlet JJs and for low frequencies in the case of triplet JJs. At low temperatures~$T$, the real part of the admittance~$Y^{\prime}(\Omega)$ in singlet JJs starts to increase from zero at ${\hbar \Omega \geq \Delta_{\text{sg}}}$, but at ${T \geq \Delta_{\text{sg}}}$, it has a peak at low frequencies the magnitude of which is determined by inelastic processes. The subgap~$\Delta_{\text{sg}}$ depends on transparencies of the S/N~interfaces and on the phase difference~$2 \chi_{0}$. The low-frequency peak in~$Y^{\prime}(\Omega)$ in triplet JJs disappears.
\end{abstract}

\date{\today}
\maketitle

\affiliation{Theoretische Physik III, Ruhr-Universit\"{a}t Bochum, D-44780 Bochum, Germany}

\affiliation{Theoretische Physik III, Ruhr-Universit\"{a}t Bochum, D-44780 Bochum, Germany}

\section{Introduction}

The frequency dependence of the admittance~$Y(\Omega)$ for uniform superconductors has been calculated long ago in the well known papers by Mattis and Bardeen and by Abrikosov, Gor'kov, Khalatnikov.\cite{Mattis_Bardeen_1958,Abrikosov_Gorkov_Khalatnikov_1958,AbrGor58} It has been shown that the real part of the admittance~$Y^{\prime}(\Omega)$ at zero temperature is zero unless the frequency~$\Omega$ does not exceed~$2 \Delta / \hbar$. This natural result has been confirmed experimentally.\cite{GloverTinkham56,GinsbergTinkham60,tinkham2004introduction} The admittance of weakly inhomogeneous superconductors has been determined by Larkin and Ovchinnikov.~\cite{LO72} These authors have also calculated the admittance of a strongly inhomogeneous superconductor with a sufficiently strong Zeeman interaction in the so-called Fulde-Ferrel-Larkin-Ovchinnikov state.\cite{Larkin_Ovchinnikov_1965}

Strongly inhomogeneous superconductivity is realized also in S/N~structures or in Josephson weak links of different kinds, S/N/S, S/c/S etc., where~N is a normal film and~c means a constriction. Superconducting correlations are induced in the~N region due to proximity effect so that a subgap~$\Delta_{\text{sg}}$ (${\Delta_{\text{sg}} < \Delta}$) may arise in the N~film.\cite{McMillan_1968,GolKupr89} The admittance~$Y(\Omega)$ for the S/N/N$_{\text{res}}$ structure was calculated in Ref.~\onlinecite{Volkov199321}, where N$_{\text{res}}$ means a bulk normal metal (reservoir) or a thick normal film attached to a thinner N~film or wire. A peak in~$Y^{\prime}(\Omega)$ was shown to exist at a frequency corresponding to a subgap~$\Delta_{\text{sg}}$ in the N~film.

More attention was paid to the study of ac properties of superconducting weak links (see reviews Refs.~\onlinecite{Likharev,Barone_Paterno_1982,Kulik72}). The interest in this study rose in recent years due to rapid progress in experimental techniques and possible applications of the Josephson junctions.\cite{Goltsman_et_al_2001,Day_et_al_2003,Janssen_et_al_2013,Clarke_Wilhelm_2008} The formula for the admittance of a short S/c/S (or S/N/S) weak link was derived in Ref.~\onlinecite{AVZ79} and analyzed in more detail in our recent paper~\onlinecite{MV17} (see also Ref.~\onlinecite{Kos_Nigg_Glazman_2013} where another approach of calculations has been used). The admittance of long S/N/S junctions has been calculated in Refs.~\onlinecite{Virtanen_et_al_2011,Tikhonov_Feigelman_2015}. In all these papers, an anomalous enhancement of the real part of admittance~$Y^{\prime}(\Omega)$ at low frequencies~$\Omega$ has been obtained.\cite{AVZ79,Kos_Nigg_Glazman_2013,Virtanen_et_al_2011,Tikhonov_Feigelman_2015,MV17} The enhancement is caused by quasiparticles with energies in the interval ${\Delta_{\text{sg}} < \epsilon < \Delta}$, where the subgap~$\Delta_{\text{sg}}$ depends on the dc phase difference~$2 \chi_{0}$.

Interestingly, collective modes may lead to peculiarities in the admittance~$Y^{\prime}(\Omega)$ of a uniform superconductor or weak links under certain conditions. For example, the amplitude mode results in a peak at ${\Omega \simeq 2 \Delta}$ in admittance of a current carrying superconductor,\cite{MVEPRL17} and the phase mode, or the so-called Carlson-Goldman mode,\cite{Carlson_Goldman_1975,Schmid_Schoen_1975,ArtVolkov75,Schoen84_a,AVZcoll78,ArtVolkovRev80} leads to some features in the $I$\nobreakdash-$V$ characteristics and in the dependence~$Y(\Omega)$.\cite{AVZcoll78,ArtVolkovRev80}

On the other hand, in the last decade a great deal of attention was paid to the study of triplet, odd-frequency (TOF) superconductivity,\footnote{The TOF component is often called the Berezinskii's type of superconductivity.\cite{Berezinskii_1974} In our opinion, this term is not suitable for the description of the TOF superconducting correlations that arise in S/F~systems. Berezinskii has suggested a new type of superconductivity caused by a special (hypothetical) type of coupling which is described by a frequency-dependent potential. On the other hand, the TOF correlations arise even in conventional BCS superconductors in the presence of a magnetic field acting on spins of electrons (Zeeman's splitting). In this case one can hardly speak about a new type of superconductivity.} which arises in S/F~structures (see reviews Refs.~\onlinecite{GolubovRMP04,BVErmp,BuzdinRMP,Eschrig_Ph_Today,BlamireRev14,Eschrig_Reports_2015,Linder_Robinson_2015} and references therein). It was shown\cite{Bergeret_Volkov_Efetov_2001} that the proximity effect induces triplet Cooper pairs in S/F~structures with an inhomogeneous magnetization~$M(x)$ in the ferromagnet~F.\footnote{Similar ideas were suggested a little later in Ref.~\onlinecite{Kad01}, where this effect was discussed qualitatively.} These pairs penetrate into the ferromagnet over a relatively large distance~$l_{T}$ which may be of the order ${l_{T} \sim \sqrt{D/T}}$, that is, much larger than than the length ${l_{h} \sim \sqrt{D/h}}$ of the condensate penetration into a homogeneous ferromagnet with an exchange field~$h$ ($D$~is the diffusion coefficient and $T$---the temperature). The long penetration of the condensate is caused by the triplet Cooper pairs with the total spin orientation parallel to the magnetization vector~$\mathbf{M}$ in the ferromagnet~F. In the case of a magnetically homogeneous ferromagnet triplet Cooper pairs penetrating into~F from a singlet superconductor~S have the total spin oriented perpendicular to the vector~$\mathbf{M}$ and penetrate the ferromagnet over a rather short distance~$\sim l_{h}$.

The long range triplet superconducting correlations may provide, in particular, the Josephson coupling in S/F/S junctions with a relatively thick ferromagnetic film~F. The appearance of these long range triplet component has been proved mainly by observing the dc Josephson effect in JJs of different kinds with a ferromagnetic layer(s).\cite{Keizer06,Petrashov06,Aarts10,Aarts11,Aarts12,Birge10,Birge12,Zabel10,Blamire10,BlamirePRL10,Petrashov11,Moor_Volkov_Efetov_2015_d,Richard_Buzdin_Houzet_Meyer_2015} Different aspects of the stationary Josephson effect in Josephson junctions of various types were analyzed in many theoretical papers (see Refs.~\onlinecite{BVE03,Eschrig03,Fominov05,Buzdin06,Braude07,Tanaka07,Buzdin07,Zaikin08,Eschrig08,Grein_et_al_2009,Buzdin09,Radovic11,Radovic11a,Bobkov11,Pugach_Buzdin_2012,Halterman14,Linder13,LinderKul14,Linder14,Hikino_Yunoki_2015,Valls15} as well as references in recent review articles Refs.~\onlinecite{BlamireRev14,Eschrig_Reports_2015,Linder_Robinson_2015}). On the other hand, dynamics of the triplet component was studied only in uniform superconductors,\cite{FominovPRL11,TanakaPRB12} but not in Josephson junctions.

In this paper, we calculate and analyze the admittance~$Y(\Omega)$ of the JJs of two types, S/N/S and S$_{\text{m}}$/N/S$_{\text{m}}$ (or S$_{\text{m}}$/F/S$_{\text{m}}$). In the latter case it doesn't matter whether superconductors~S$_{\text{m}}$ are connected by a normal or ferromagnetic wire. It is merely important that the magnetization vector~$\mathbf{M}$ is parallel to the filters axis, i.e., ${\mathbf{M} || \mathbf{e}_{z}}$. The Josephson coupling is provided by the singlet component in the JJs of the first type and by the triplet odd-frequency component in the JJs of the second type. We consider the case of short JJs with the distance between the superconducting reservoirs shorter than the coherence length ${\xi_{\text{S}} \sim \{ \sqrt{D/T}, \sqrt{D/\Delta} \}}$. Unlike the case studied in Refs.~\onlinecite{AVZ79,MV17}, where the S/N~interface resistance~$R_{\text{if}}$ was assumed to be negligible compared with the resistance~$R_{\text{N}}$ of the~N region, we consider here the opposite case, i.e., ${R_{\text{if}} \gg R_{\text{N}}}$. As in the case considered in Refs.~\onlinecite{AVZ79,MV17}, the real part of the admittance~$Y^{\prime}(\Omega)$ in singlet JJs has a maximum at low~$\Omega$ if the temperature is not too low, ${T \gtrsim \Delta_{\text{sg}}}$. The low frequency behavior of~$Y^{\prime}(\Omega)$ is described approximately by the expression ${Y^{\prime}(\Omega) \sim \big[ 2 \gamma_{\text{N}} / (4 \gamma_{\text{N}}^{2} + \Omega^{2}) \big] \exp (-\Delta_{\text{sg}} / T)}$. Although general formulas for the current obtained in this paper for the singlet weak links, Eqs.~(\ref{B17}),~(\ref{B18}),~and(\ref{B19}), look similar to those in our previous publication~Ref.~\onlinecite{MV17}, they differ essentially because the integrands in corresponding equations are different.\footnote{Note misprints in Ref.~\onlinecite{MV17}. The denominators in Eqs.~(60)~and~(61) should be the same as in Eq.~(B6). The sign of~$I_{\Omega}^{\text{an}}$ must be changed.} In the case of triplet coupling, a little enhancement of~$Y^{\prime}(\Omega)$ remains only if a characteristic exchange field~$h$ is small compared with a subgap~$\Delta_{\text{sg}}$. Otherwise no enhancement in~$Y^{\prime}(\Omega)$ appears.

\begin{figure}
\includegraphics[width=1.0\columnwidth]{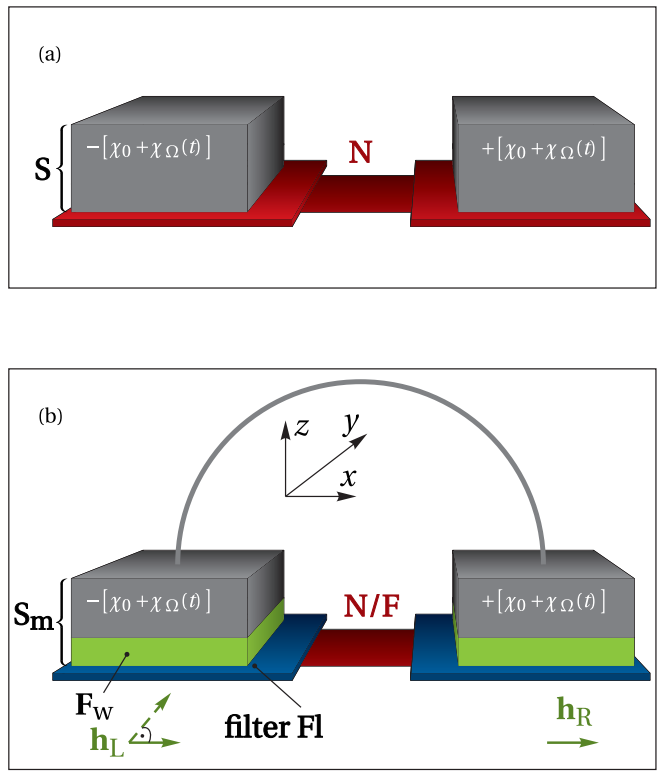}
\caption{(Color online.) Schematic representation of Josephson junctions under consideration. (a)~JJ with singlet coupling; (b)~JJ with triplet coupling. The superconductors S$_{\text{m}}$ consist of conventional singlet superconductors covered by thin layers~F$_{\text{w}}$ of weak ferromagnets so that triplet Cooper pairs with spin polarizations in $x$-$z$ or $y$-$z$ planes penetrate due to proximity effect into the F$_{\text{w}}$~layer. The filters~Fl let pass Cooper pairs with spins parallel or antiparallel to the $z$~axis. If the exchange fields~$h_{\text{R},\text{L}}$ at the right and at the left have different directions, a spontaneous current arises which can flow through the shown loop.}
\label{fig:Scheme}
\end{figure}

The plan of the paper is as follows. In the Section~\ref{sec:basic_eqns}, we present the main equations. In Section~\ref{sec:singlet_coupling}, we calculate the critical Josephson current~$I_{\text{c}}$ for an S/N/S Josephson weak link with a coupling via the singlet component. We find also a response of this JJ to a small ac phase variation and present an expression for the admittance of the junction~$Y(\Omega)$. In Section~\ref{sec:triplet_coupling}, we obtain and analyze the current~$I_{\text{c}}$ and the admittance~$Y(\Omega)$ in JJs of the S$_{\text{m}}$/N/S$_{\text{m}}$ or S$_{\text{m}}$/F/S$_{\text{m}}$ types, where~S$_{\text{m}}$ is a ``magnetic'' superconductor serving as a source of fully polarized triplet Cooper pairs [see Fig.~\ref{fig:Scheme}~(b)]. In Section~\ref{sec:conclusions}, the obtained results are discussed.

\section{Basic Equations}
\label{sec:basic_eqns}

We consider JJs of two types shown in Fig.~\ref{fig:Scheme}. The JJ or weak links in Fig.~\ref{fig:Scheme}~(a) consist of two bulk BCS superconductors~S connected by a normal wire (or film). The JJ depicted in Fig.~\ref{fig:Scheme}~(b) consists of two bulk ``magnetic'' superconductors~S$_{\text{m}}$ connected by a normal (or ferromagnetic) wire. The superconductors S$_{\text{m}}$ are formed by a BCS superconductor covered by a thin film of a weak ferromagnet~F with an exchange field~$\mathbf{h}$. We assume that at the~S$_{\text{m}}$/N interfaces there are spin filters which let pass electrons with only one spin direction (${\mathbf{s}||\mathbf{z}}$). The exchange field~$\mathbf{h}$ is supposed to be perpendicular to the~$z$~axis so that triplet Cooper pairs, which appear in the~F film, have a component along the $z$~axis and therefore penetrate through the filters. The singlet Cooper pairs do not pass through the filters. We consider the diffusive case assuming that the mean free path is shorter than the coherence length ${\xi_{\text{S}} \simeq \sqrt{D / T_{\text{c}}}}$.

In order to find the current~$I$ through the system, we need to determine quasiclassical Green's functions. We employ the same basic equations for the quasiclassical matrix Green's functions~$\check{g}$ as in Ref.~\onlinecite{MV17}. In the considered one-dimensional geometry and in the diffusive limit they obey a generalized Usadel equation of the form\cite{Kopnin,LO,RammerSmith,BelzigRev,ArtVolkovRev80}
\begin{equation}
-i D \partial_{x}(\check{g} \partial_{x} \check{g}) + i( \check{\tau}_{3} \partial_{t} \check{g} + \partial_{t^{\prime }} \check{g} \check{\tau}_{3}) + [\check{\Sigma} \,, \check{g}] = V(t) \check{g} - \check{g} V (t^{\prime}) \,,
\label{1}
\end{equation}
where~$D$ is the diffusion coefficient, the matrix~$\check{\Sigma}$ describes damping, and~$V$ is the electric potential. Equation~(\ref{1}) describes the Green's function in the N film and is complemented by the boundary condition\cite{Bergeret12b,EschrigBC13,EschrigBC15}
\begin{equation}
\check{g} \partial_{x} \check{g} = \pm \varkappa_{\text{L},\text{R}}[\check{g} \,, \mathrm{\check{\Gamma}} \cdot \mathrm{\check{G}} \mathrm{\check{\Gamma}}]|_{\pm L} \,,  \label{2}
\end{equation}
where ${\varkappa_{\text{L},\text{R}} = 1/R_{\Box \text{L},\text{R}} \sigma}$, $R_{\Box \text{L},\text{R}}$ is the interface resistance of the left (right) interface per unit area, and~$\sigma$ is the conductivity of the N~metal. Note that this boundary condition insures the continuity of the current across the S$_{\text{m}}$/N or S$_{\text{m}}$/F interfaces. The matrix ${\mathrm{\check{\Gamma}} = \mathrm{diag} \big\{\mathrm{\hat{\Gamma},\hat{\Gamma}}\big\}}$ describes the action of the filters. If the filters allow to pass only electrons with spins aligned parallel to the $z$~axis, then ${\hat{\mathrm{\Gamma}} = (\mathcal{T} \hat{1} + \mathcal{U} \hat{\mathrm{X}}_{33}) / \sqrt{2}}$ with ${\mathcal{T} = \pm \mathcal{U}}$ and and ${\mathrm{\hat{X}}_{ij} = \hat{\tau}_{i} \cdot \hat{\sigma}_{j}}$ for ${i,j = 0,1,2,3}$, where~$\tau_i$ and~$\sigma_j$ are Pauli matrices operating in Gor'kov-Nambu, respectively, spin spaces ($\tau_0$ and~$\sigma_0$ are $2 \times 2$~unity matrices). The probability for an electron with spin up (down) to pass into the N~wire is ${\mathcal{T}_{\uparrow,\downarrow} \propto \mathcal{T} \pm \mathcal{U}}$. We set ${\mathcal{U} = s \mathcal{T}}$ with ${s = \pm 1}$ and the coefficients~$\mathcal{T}$ and~$\mathcal{U}$ are normalized so that ${\mathcal{T} = |\mathcal{U}| = 1}$. The sign of the factor~$s$ determines the orientation of the spins of electrons (parallel or antiparallel to the vector~$\mathbf{e}_{z}$) passing through the filters. In the case of spin-inactive interfaces~${\hat{\mathrm{\Gamma}} = \hat{1}}$. We have to solve the Usadel equation, Eq.~(\ref{1}), for the Green's function~$\check{g}$ in the N~film with the boundary condition Eq.~(\ref{2}) that connects the Green's function~$\check{g}$ in the N~film with the known Green's functions~$\check{G}$ in the reservoirs (please, note the use of lowercase and capital letters to denote the Green's functions in different systems). We will neglect the inverse proximity effect on the superconducting reservoirs (see Appendix~\ref{app:inv_prox_eff}).

The matrix Green's function~$\check{g}$ consists of the retarded (advanced) Green's functions~$\hat{g}^{R(A)}$ (diagonal elements) and the Keldysh function~$\hat{g}$ (the off-diagonal~$\check{g}_{12}$ element) and obeys the normalization condition
\begin{equation}
\check{g} \cdot \check{g} = \check{1} \,. \label{1b}
\end{equation}

The Thouless energy ${E_{\text{Th}} = D / L^{2}}$ is assumed to be much larger than~$\Delta$ and~$V$. This assumption means that the function~$\check{g}(x)$ is almost constant in space. Integrating Eq.~(\ref{1}) over~$x$ and taking into account the boundary condition Eq.~(\ref{2}), we obtain for the Fourier component~$\check{g}(\epsilon,\epsilon^{\prime })$
\begin{equation}
\epsilon \mathrm{\check{X}}_{30} \cdot \check{g} - \check{g} \cdot \mathrm{\check{X}}_{30} \epsilon^{\prime} = i E_{0} \big[\check{g}\,,\check{G}\big] \,,  \label{3}
\end{equation}
where ${\mathrm{\check{G}} \equiv (\mathrm{\check{G}}_{\text{R}} + \mathrm{\check{G}}_{\text{L}}) / 2}$, ${\mathrm{\check{X}}_{30} = \mathrm{diag}\big\{\mathrm{\hat{X}}_{30} \,, \mathrm{\hat{X}}_{30}\big\}}$, and ${E_{0} = D \varkappa / L}$. For simplicity we assume that the interface resistances are equal, ${\varkappa_{\text{R}} = \varkappa_{\text{L}} \equiv \varkappa}$. The matrices~$\mathrm{\check{G}}$ describe electrons passing through the filter,
\begin{equation}
\check{G} = \mathrm{\check{\Gamma}} \cdot \mathrm{\check{G}} \cdot \mathrm{\check{\Gamma}} \,. \label{3'}
\end{equation}
The matrices~$\check{G}$ and~$\mathrm{\check{G}}$ consist of the retarded (advanced) ($\hat{G}^{R(A)}$) and Keldysh ($\hat{G}$) Green's functions in the reservoirs,
\begin{equation}
\check{G}= \begin{pmatrix}
\hat{G}^{R} & \hat{G} \\
0  & \hat{G}^{A}
\end{pmatrix} \,.  \label{3a}
\end{equation}
The retarded (advanced) Green's functions $\hat{G}^{R(A)}$ are given by
\begin{equation}
\mathrm{\check{G}}^{R(A)}(t,t^{\prime}) = \mathrm{\hat{S}}(t) \big[ G(t-t^{\prime}) \mathrm{\hat{X}}_{30} + F(t-t^{\prime}) \mathrm{\hat{X}}_{10} \big]^{R(A)} \mathrm{\hat{S}}^{\dagger}(t^{\prime}) \,, \label{3b}
\end{equation}
where ${\mathrm{\hat{S}}(t) = \exp (\mathrm{\hat{X}}_{30} i \chi(t)/2)}$ with the phase of the order parameter in the right, respectively, left reservoir ${\chi_{\text{R},\text{L}}(t) = \pm \big[\chi_{0} + \chi_{\Omega}(t)\big]_{\text{R},\text{L}}}$. We set ${\chi_{\text{R}}(t) = -\chi_{\text{L}}(t) \equiv \chi(t)}$. If the reservoirs are BCS singlet superconductors, the Fourier components of the functions~$G^{R(A)}(t-t^{\prime})$ and~$F^{R(A)}(t-t^{\prime})$ are
\begin{align}
G_{0}^{R(A)}(\epsilon) &= (\epsilon \pm i\gamma) / \zeta^{R(A)} \,, \\
F_{0}^{R(A)}(\epsilon) &= \Delta / \zeta^{R(A)} \,, \label{3c}
\end{align}
with ${\zeta^{R(A)} = \sqrt{(\epsilon \pm i \gamma)^{2} - \Delta^{2}}}$ and the damping rate in the superconducting reservoirs~$\gamma$.

In the case of S$_{\text{m}}$ superconductors, the Green's functions~$\hat{G}(\omega)$ have a more complicated structure.\cite{Moor_Volkov_Efetov_2016_a} In a static case (${\chi = \chi_{0} = \text{const}}$), in the Matsubara representation (${\epsilon = i \omega \equiv i \pi T(2n+1)}$) they have the form
\begin{equation}
\hat{G}(\omega) = \mathrm{\hat{S}}_{0} \big[ G_{\omega h} \big( \mathrm{\hat{X}}_{30} + s \mathrm{\hat{X}}_{03} \big) + F_{\omega h} \mathrm{\hat{X}}_{\perp} \big] \mathrm{\hat{S}}_{0}^{\dagger} \,. \label{4}
\end{equation}%
Here, ${\mathrm{\hat{S}}_{0} = \exp (\mathrm{\hat{X}}_{30} i \chi_{0}/2)}$ and
\begin{align}
G_{\omega h} &= \frac{1}{2} \Big(\frac{\omega + i h}{\zeta_{\omega +}} + \frac{\omega -i h}{\zeta_{\omega -}} \Big) \equiv \mathrm{Re}\Big( \frac{\omega + i h}{\zeta_{\omega +}} \Big) \,, \label{5} \\
F_{\omega h} &= \frac{\Delta}{2} \Big(\frac{1}{\zeta_{\omega +}} - \frac{1}{\zeta_{\omega -}}\Big) \equiv i \Delta \mathrm{Im} \Big(\frac{1}{\zeta_{\omega +}} \Big) \,, \label{5'}
\end{align}
with ${\zeta_{\omega \pm} = \sqrt{(\omega \pm i h)^{2} + \Delta^{2}}}$. The factor ${s = \pm 1}$ characterizes the direction of the triplet Cooper pairs respective to the z~axis. The form of the matrix~$\mathrm{\hat{X}}_{\perp}$ depends on the direction of the exchange field~$\mathbf{h}$ ($\mathbf{h}||\mathbf{e}_{x}$ or $\mathbf{h}||\mathbf{e}_{y}$),
\begin{equation}
\mathrm{\hat{X}}_{\perp} = \frac{1}{\sqrt{2}}
\begin{cases}
\mathrm{\hat{X}}_{11} - s \mathrm{\hat{X}}_{22} \,, & \mathbf{h} || \mathbf{e}_{x} \,, \\
\mathrm{\hat{X}}_{12} + s_{\text{L}} \mathrm{\hat{X}}_{21} \,, & \mathbf{h} || \mathbf{e}_{y} \,.
\end{cases}  \label{st3}
\end{equation}
We clarify the origin of matrices in this equation. If the exchange field~$h$ is parallel to the $z$~axis ($h || e_{z}$), then, the part of the´triplet condensate Green's function is ${\hat{F}_{\omega h} = F_{\omega h} \hat{X}_{13}}$, where the function~$F_{\omega h}$ is defined in Eq.~(\ref{5'}). Rotation around the $x$,~respectively,~$y$~axis with the aid of the rotation matrix ${\hat{U} = \cos \alpha / 2 + i \hat{X}_{01(2)} \sin \alpha / 2}$ with ${\alpha = \pi /2}$ transforms the function ${\hat{F}_{\omega h} \to F_{\omega h} \hat{X}_{12}}$ or ${\hat{F}_{\omega h} \to F_{\omega h} \hat{X}_{11}}$. Action of the spin-filter transforms these functions into ${\hat{\Gamma}\hat{F}_{\omega h} \hat{\Gamma} \to F_{\omega h} (\hat{X}_{12} + s \hat{X}_{21})}$ or ${\hat{\Gamma} \hat{F}_{\omega h} \hat{\Gamma} \to F_{\omega h} (\hat{X}_{11} - s \hat{X}_{22})}$.

We need to find the current~$I$ given by the expression (below this equation will be written in more detailed form; see Appendix~\ref{app:ac_current_formulas})
\begin{align}
I &= (16 \kappa R_{\Box} e)^{-1} \int d \bar{\epsilon} \, \big\{ (\check{g} \partial_{x} \check{g})^{K} \big\}_{30} \notag \\
&= (16 R_{\Box }e)^{-1} \int d \bar{\epsilon} \, \big\{ \big[ \check{g} \,, \mathrm{\check{\Gamma} \check{G}} \mathrm{\check{\Gamma}} \big]^{K} \big\}_{30} \label{6}
\end{align}
as a response to a small periodic variation of the phase ${\chi_{\Omega}(t) = \chi_{\Omega} \cos (\Omega t)}$ in the presence of a constant phase difference~$\chi_{0}$. Here, ${\bar{\epsilon} = (\epsilon + \epsilon^{\prime}) / 2}$ and we introduced the notation ${\big\{ ( \check{g} \partial_{x} \check{g} )^{K} \big\}_{ij} \equiv \mathrm{Tr} \big\{\mathrm{\hat{X}}_{ij} (\check{g} \partial_{x} \check{g})^{K} \big\}/4}$. Therefore, we have to solve Eq.~(\ref{3}) and to find the function~$\check{g}$. First, we consider the case of singlet superconductors.

Unlike a similar system with perfectly penetrable interfaces considered in Ref.~\cite{AVZ79,MV17}, in our case there are barriers at the S/N interfaces so that the interfaces have a finite resistance ${R_{\text{if}} \equiv R_{\Box}}$. Moreover, we assume that the interface resistance is much larger than the resistance of the N~layer, ${R_{L} = 2 L / \sigma}$, i. e., the inequality
\begin{equation}
R_{\Box} \gg R_{L} \label{6a}
\end{equation}
is fulfilled.

\section{Singlet Coupling}
\label{sec:singlet_coupling}

In this Section, we consider an S/N/S junction in the absence of spin filters at the interfaces, i.e., we set ${\mathrm{\check{\Gamma}} = \mathrm{\check{1}}}$. This means that the matrices~$\mathrm{\hat{G}}$ and~$\hat{G}$ coincide. Therefore, the Josephson coupling is realized through singlet Cooper pairs penetrating the N~region due to proximity effect. First, we analyze the stationary case.

\subsection{Stationary case}

In this case, it is convenient to use the Matsubara representation of the
Green's functions. The matrices $\big[ \hat{G}_{\omega} \big]_{\text{R},\text{L}}$ read
\begin{equation}
\big[ \hat{G}_{\omega} \big]_{\text{R},\text{L}} = \big[ G_{\omega} \mathrm{\hat{X}}_{30} + F_{\omega} \exp (\pm i \chi_{0} \mathrm{\hat{X}}_{30} ) \cdot \mathrm{\hat{X}}_{10} \big]_{\text{R},\text{L}} \,. \label{B1}
\end{equation}
The function~$G_{\omega}$ is equal to~$G_{\omega h}$ at ${h = 0}$ [see Eq.~(\ref{5})] and ${F_{\omega} = ( \Delta / \omega ) G_{\omega}}$.

From Eq.~(\ref{3}) we obtain an equation for the stationary Green's function $\hat{g}_{\omega}$. One can write this equation in the form
\begin{equation}
\big[ \mathrm{\hat{M}}_{\omega} \,, \hat{g}_{\omega} \big] = 0 \,, \label{B2}
\end{equation}
where ${\mathrm{\hat{M}}_{\omega} = \tilde{\omega} \mathrm{\hat{X}}_{30} + \tilde{\Delta}_{\omega} \mathrm{\hat{X}}_{10}}$ and ${\tilde{\omega} = \omega ( 1 + E_{0} / \zeta_{\omega} )}$, ${\tilde{\Delta}_{\omega} = [ E_{0} \Delta / \zeta_{\omega} ] \cos \chi_{0}}$, ${\zeta_{\omega} = \sqrt{\omega^{2} + \Delta^{2}}}$, ${\tilde{\zeta}_{\omega} = \sqrt{\tilde{\omega}^{2} + \tilde{\Delta}_{\omega}^{2}}}$. A solution satisfying the normalization condition, Eq.~(\ref{1b}), is
\begin{equation}
\hat{g}_{\omega} = \frac{\mathrm{\hat{M}}_{\omega}}{\tilde{\zeta}_{\omega}} \,.
\label{B5}
\end{equation}

The retarded function ${\hat{g}^{R}(\epsilon) = g_{\epsilon}^{R} \mathrm{\hat{X}}_{30} + f_{\epsilon}^{R} \mathrm{\hat{X}}_{10}}$, which is equal to~$\hat{g}_{\omega}$ at ${\omega = -i \epsilon}$, determines the density of states (DOS) ${\nu(\epsilon) = \mathrm{Re} ( g_{\epsilon} )}$. One can easily obtain a transparent formula for $\nu(\epsilon)$ in the limiting cases of large and small~$E_{0}$.

For large~$E_{0}$ (${E_{0} \gg \Delta}$) we have ${\tilde{\epsilon} \simeq \epsilon_{0} / \zeta_{0}^{R}}$ and ${\tilde{\Delta} \simeq \big( \zeta_{0}^{R} \big)^{-1} \Delta_{0} \cos \chi_{0}}$ with ${\zeta_{0}^{R} = \sqrt{(\epsilon + i \gamma_{\text{N}})^{2} - \Delta_{0}^{2} \cos^{2} \chi_{0}}}$. We obtain the standard formula for a BCS superconductor ${\nu(\epsilon) = \mathrm{Re}\big( \epsilon / \sqrt{\epsilon^{2} - (\Delta \cos \chi_{0})^{2}} \big)}$ with a gap in the quasiparticle spectrum~$\Delta |\cos \chi_{0}|$. In the opposite limit of small~$E_{0}$ (${E_{0} \ll \Delta}$), we obtain ${\nu(\epsilon) = \mathrm{Re}\big( \epsilon / \sqrt{\epsilon^{2} - (E_{0} \cos \chi_{0})^{2}}\big)}$, that is, the energy gap or subgap ${\Delta_{\text{sg}} = E_{0} \cos \chi_{0}}$ is much smaller than~$\Delta$.\cite{McMillan_1968}

One can easily calculate the dc Josephson current~$I_{\text{J}}$ using Eqs.~(\ref{6}) and~(\ref{B5}). In the considered static case, the expression for the current can be written in the form
\begin{equation}
I_{\text{J}} = \frac{i \pi T}{2 e R_{\Box}} \sum_{\omega \geq 0} \big\{ [\hat{g}_{\omega} \,, \hat{G}_{\omega \text{R}}] \big\}_{30} \,. \label{B6}
\end{equation}
Here, we used the form of the Keldysh function ${\hat{g} = (\hat{g}^{R} - \hat{g}^{A}) \tanh (\epsilon \beta)}$ [in this case, ${\epsilon = \epsilon^{\prime} = \bar{\epsilon} \equiv (\epsilon + \epsilon^{\prime})/2}$] and transformed the integral in Eq.~(\ref{6}) into the sum over poles of~$\tanh (\epsilon \beta)$. Substituting Eqs.~(\ref{B1}) and~(\ref{B5}) into this equation, we find a standard relation for the Josephson current,\cite{Likharev,Barone_Paterno_1982,Kulik72}
\begin{equation}
I_{\text{J}} = I_{\text{c}} \sin (2 \chi_{0}) \,, \label{B7a}
\end{equation}
where the critical current of the considered JJ with the Josephson coupling through the singlet condensate equals
\begin{equation}
I_{\text{cS}} = \frac{2 \pi T}{e R_{\Box}} E_{0} \Delta^{2} \sum_{\omega \geq 0} \frac{1}{\tilde{\zeta}_{\omega} \zeta_{\omega}^{2}} \,. \label{B7b}
\end{equation}

Now we turn to the response of the singlet JJ to an ac voltage~$V_{\Omega}$.

\subsection{Non-stationary case}

We have to find a response~$\delta \check{g}$ to a phase variation ${\chi_{\Omega}(t) = \chi_{\Omega}(t)_{\text{R}} = - \chi_{\Omega}(t)_{\text{L}}}$ or to an ac voltage~$V_{\Omega}(t)$ applied to the considered JJ and coupled to~$\chi_{\Omega}(t)$ via the Josephson relation
\begin{equation}
2 e V_{\Omega}(t) = \hbar \partial \chi_{\Omega}(t) / \partial t \,. \label{B8b}
\end{equation}

In order to determine a variation of the Green's function in the N~region~$\delta \hat{g}^{R(A)}$, we linearize Eq.~(\ref{3}) and take into account that in case of harmonic variation of the phase, ${\chi_{\Omega}(t) = \chi_{\Omega} \exp (-i \Omega t)}$, the functions~$\delta \hat{g}^{R(A)}(\epsilon,\epsilon^{\prime})$ can be represented in the form ${\delta \hat{g}^{R(A)}(\epsilon,\epsilon^{\prime}) = \delta \hat{g}^{R(A)} 2 \pi \delta (\epsilon - \epsilon^{\prime} - \Omega)}$. For the functions~$\delta \hat{g}^{R(A)}$ we obtain
\begin{equation}
\big[ \hat{M}(\epsilon_{+}) \cdot \delta \hat{g} - \delta \hat{g} \cdot \hat{M}(\epsilon_{-}) \big]^{R(A)} = i E_{0} \big\{ \hat{g}_{0+} \cdot \delta \hat{G} - \delta \hat{G} \cdot \hat{g}_{0-} \big\}^{R(A)} \,, \label{B11}
\end{equation}
where ${\hat{M}^{R(A)}(\epsilon) = \big[ \epsilon \mathrm{\hat{X}}_{30} + i E_{B} \hat{G}_{0} \big]^{R(A)}}$ and ${\epsilon_{\pm}^{R} = \bar{\epsilon} \pm \Omega / 2 + i \gamma_{\text{N}}}$ (we take into account a damping rate $\gamma_{\text{N}}$ in the normal metal~N). One can represent the left-hand side in the form
\begin{equation}
\big[ \tilde{\zeta}_{+} \hat{g}_{0+} \cdot \delta \hat{g} - \delta \hat{g} \cdot \hat{g}_{0-} \tilde{\zeta}_{-} \big]^{R(A)} = i E_{0} \big\{ \hat{g}_{0+} \cdot \delta \hat{G} - \delta \hat{G} \cdot \hat{g}_{0-} \big\}^{R(A)} \,, \label{B12}
\end{equation}
where the functions ${\hat{g}_{0 \pm} = \hat{g}_{0}(\bar{\epsilon} \pm \Omega / 2)}$ and~$\delta \hat{G}^{R(A)}$ are defined in Eqs.~(\ref{3Aa})\nobreakdash--(\ref{3Ac}) (see Appendix~\ref{app:singlet}). The normalization condition Eq.~(\ref{1b}) yields
\begin{align}
(\hat{g}_{0 \pm} \cdot \hat{g}_{0 \pm})^{R(A)} &= \hat{1} \,, \label{B13} \\
\big[ \hat{g}_{0+} \cdot \delta \hat{g} + \delta \hat{g} \cdot \hat{g}_{0-} \big]^{R(A)} &= 0 \,.
\end{align}
Using these equations, we obtain a solution of Eq.~(\ref{B12}),
\begin{equation}
\delta \hat{g}^{R(A)} = \Big[ \frac{i E_{0}}{\tilde{\zeta}_{+} + \tilde{\zeta}_{-}} \big[ \delta \hat{G} - \hat{g}_{0+} \cdot \delta \hat{G} \cdot \hat{g}_{0-} \big] \Big]^{R(A)} \,.
\label{B14}
\end{equation}

The Keldysh matrix~$\delta \hat{g}$ is represented as a sum of regular and anomalous parts, $\delta \hat{g}^{\text{reg}}$, $\delta \hat{g}^{\text{an}}$,\cite{G-Eliash,ArtVolkovRev80}
\begin{equation}
\delta \hat{g} = \delta \hat{g}^{\text{reg}} + n(\bar{\epsilon}) \hat{g}^{\text{an}} \,, \label{B15}
\end{equation}
where ${n(\bar{\epsilon},\Omega) = \big[ \tanh(\epsilon_{+} \beta) - \tanh(\epsilon_{-} \beta) \big]}$ and
\begin{equation}
\delta \hat{g}^{\text{reg}} = \delta \hat{g}^{R} \tanh (\epsilon_{-} \beta) - \tanh(\epsilon_{+} \beta) \delta \hat{g}^{A} \,. \label{B15a}
\end{equation}%
The anomalous function can be found by using the same simple procedure as in
Refs.~\cite{ArtVolkovRev80,MV17}. We find
\begin{equation}
\hat{g}^{\text{an}} = \frac{i E_{0}}{\tilde{\zeta}_{+}^{R} + \tilde{\zeta}_{-}^{A}} \big[ \delta \hat{G}^{\text{an}} - \hat{g}_{0+}^{R} \cdot \delta \hat{G}^{\text{an}} \cdot \hat{g}_{0-}^{A} \big] \,. \label{B16}
\end{equation}
The expression for~$\hat{G}^{\text{an}}$ is provided in the Appendix~\ref{app:singlet}, Eq.~(\ref{7A}).

Knowing the variations of the Green's functions, $\delta \hat{g}^{R(A)}$ and $\hat{g}^{\text{an}}$, we determine the variation of the current~${\delta I_{\Omega} = I_{\Omega} \exp (-i \Omega t)}$ the Fourier component of which we write as a sum of regular, $I_{\Omega}^{\text{reg}}$, and anomalous parts,~$I_{\Omega}^{\text{an}}$. The amplitude~$I_{\Omega}$ can be written as a sum of regular,~$I_{\Omega}^{\text{reg}}$, and anomalous,~$I_{\Omega}^{\text{an}}$, parts,
\begin{equation}
I_{\Omega} = I_{\Omega}^{\text{reg}} + I_{\Omega}^{\text{an}} \,. \label{B17}
\end{equation}
Here,
\begin{equation}
I_{\Omega}^{\text{reg}} = (16 R_{\Box} e)^{-1} \int d \bar{\epsilon} \, \big\{j^{R} \tanh(\epsilon_{-} \beta) - j^{A} \tanh(\epsilon_{+} \beta) \big\}_{30} \,, \label{B18}
\end{equation}
with
\begin{equation}
j^{R(A)} = \big\{\mathrm{\hat{X}}_{30} \big[ (\hat{g}_{0+} \delta \hat{G}_{\text{R}} - \delta \hat{G}_{\text{R}} \hat{g}_{0-}) - (\hat{G}_{\text{R} 0+} \delta \hat{g} - \delta \hat{g} \hat{G}_{\text{R} 0-}) \big] \big\}_{30}^{R(A)} \,,
\label{B18a}
\end{equation}
and
\begin{equation}
I_{\Omega}^{\text{an}} = (16 R_{\Box} e)^{-1} \int d \bar{\epsilon} \, n(\bar{\epsilon}) j^{\text{an}} \,, \label{B19}
\end{equation}
where the anomalous ``current''~$j^{\text{an}}$ coincides with~$j^{R}$ if the functions~$\hat{g}_{0-}^{R}$, $\delta \hat{G}_{\text{R}}^{R}$ and $G_{0-}^{R}$ in Eq.~(\ref{B18a}) are replaced by~$\hat{g}_{0-}^{A}$, $\delta \hat{G}_{\text{R}}^{\text{an}}$ and~$G_{0-}^{A}$, correspondingly.

Using Eqs.~(\ref{B15a}) and~(\ref{B16}) for the matrices in Eqs.~(\ref{B18a}), we can write~$j^{R(A)}$ and~$j^{\text{an}}$ in terms of the known Green's functions in the reservoirs. We write $j^{\text{an}}$ as a sum of two currents,
\begin{equation}
j^{\text{an}} \equiv j_{1}^{\text{an}} + j_{2}^{\text{an}} \,, \label{j_1,2}
\end{equation}
where~$j_{1,2}^{\text{an}}$ correspond to the first (second) term in Eq.~(\ref{B18a}), respectively. They can be written as follows (see Appendix~\ref{app:singlet})
\begin{widetext}
\begin{align}
j_{1}^{\text{an}} &= -i \frac{\chi_{\Omega}}{2} \big[ \big( \tilde{g}_{0+}^{R} - \tilde{g}_{0-}^{A} \big) \big( G_{+}^{R} - G_{-}^{A} \big) - \big( F_{+}^{R} + F_{-}^{A} \big) \big( \tilde{f}_{+}^{R} + \tilde{f}_{-}^{A} \big) \cos \chi_{0} \big] \,, \label{B20} \\
j_{2}^{\text{an}} &= \chi_{\Omega} \frac{E_{0}}{\tilde{\zeta}_{+}^{R} + \tilde{\zeta}_{-}^{A}} \big[1 + \tilde{g}_{0+}^{R} \tilde{g}_{0-}^{A} + \tilde{f}_{+}^{R} \tilde{f}_{-}^{A} \big] \big( F_{+}^{R} + F_{-}^{A} \big)^{2} \sin^{2} \chi_{0} \,. \label{B20a}
\end{align}
\end{widetext}
The ``currents''~$j^{R(A)}$ are given by the same formulas if all the functions of the form~$F_{-}^{A}$ ($F_{+}^{R}$) are replaced by~$F_{-}^{R}$~($F_{+}^{A}$).

One can show that the ``anomalous current'' leads to an enhancement of the conductance~$Y_{\Omega}$ at low frequencies as it takes place in S/c/S~junctions without barriers\cite{AVZ79} and in long S/N/S~junctions.\cite{Virtanen_et_al_2011,Tikhonov_Feigelman_2015} Consider the case of low frequencies (${\Omega \ll \Delta_{\text{sg}}}$). The main contribution to the current is due to the first term in Eq.~(\ref{B20}) for~$j_{1}^{\text{an}}$ and to the the ``current''~$j_{2}^{\text{an}}$. We obtain
\begin{equation}
I_{\Omega }^{\text{an}} = \frac{V_{\Omega}}{2R_{\Box}} C_{\Box} + I_{2}^{\text{an}} \,,  \label{B21}
\end{equation}
where the temperature dependent function~$C_{\Box}$ is
\begin{equation}
C_{\Box} = \int_{\Delta}^{\infty} d(\epsilon \beta) \, \frac{\epsilon \tilde{\epsilon}}{\zeta(\epsilon) \tilde{\zeta}(\epsilon) \cosh^{2}(\epsilon \beta)} \,. \label{B21'}
\end{equation}
The first term determines the admittance of the system,~$1/(2R_{\Box})$, in the normal state. The second term in Eq.~(\ref{B21}) is given by
\begin{widetext}
\begin{equation}
I_{2}^{\text{an}} = -\frac{\hbar \Omega E_{0}}{16 R_{\Box} e} \chi_{\Omega} \int d \bar{\epsilon} \, \beta \frac{1 + \tilde{g}_{0+}^{R} \tilde{g}_{0-}^{A} + \tilde{f}_{+}^{R} \tilde{f}_{-}^{A}}{\tilde{\zeta}_{+}^{R} + \tilde{\zeta}_{-}^{A}} \Big(\frac{F_{+}^{R} + F_{-}^{A}}{\cosh(\bar{\epsilon}\beta)}\Big)^2 \sin^{2} \chi_{0} \,.
\label{B22}
\end{equation}
\end{widetext}

We take into account that in the interval ${\Delta_{\text{sg}} \equiv E_{0} |\cos \chi_{0}| \leq \bar{\epsilon} \ll \Delta}$ the functions ${F_{+}^{R} = F_{-}^{A} \simeq -i}$ and ${\tilde{f}_{+}^{R} \simeq -\tilde{f}_{-}^{A} \simeq \tilde{f}^{R}}$ so that ${1 + \tilde{g}_{0+}^{R} \tilde{g}_{0-}^{A} + \tilde{f}_{+}^{R} \tilde{f}_{-}^{A} \simeq -2 (\tilde{f}^{R})^{2} = -2 \tilde{\Delta}^{2} / \tilde{\zeta}^{2}}$, and ${\tilde{\zeta}_{+}^{R} + \tilde{\zeta}_{-}^{A} \simeq (\Omega + 2 i \gamma_{\text{N}}) \bar{\epsilon} / \tilde{\zeta}}$ with ${\tilde{\zeta} = \sqrt{\bar{\epsilon}^{2} - \tilde{\Delta}^{2}}}$. Thus, we obtain
\begin{equation}
e R_{\Box} I_{2}^{\text{an}} = \frac{E_{0}^{2} \sin^{2} (2 \chi_{0})}{8 T |\cos \chi_{0}|} \frac{2 \gamma_{\text{N}} + i \Omega}{(2 \gamma_{\text{N}})^{2} + \Omega^{2}} J(\chi _{0}) \,,
\label{B23}
\end{equation}
where the integral $J(\chi _{0})$ is
\begin{align}
J(\chi _{0}) &= \int_{1}^{\infty} \frac{dx}{\cosh^{2}(x \tilde{\beta}) x \sqrt{x^{2} - 1^{2}}} \label{B24} \\
&= \begin{cases}
\pi /2 \,, & \Delta_{\text{sg}} \ll 2 T \,, \\
2 \sqrt{2 T / \Delta_{\text{sg}}} \exp (-\Delta_{\text{sg}} / T) \,, & 2 T \ll \Delta_{\text{sg}} \ll \Delta \,, \notag
\end{cases}
\end{align}
where ${\tilde{\beta} = \Delta_{\text{sg}} / 2 T}$ and ${\Delta_{\text{sg}} \equiv E_{0} |\cos \chi_{0}|}$.

\begin{figure}[tbp]
\includegraphics[width=1.0\columnwidth]{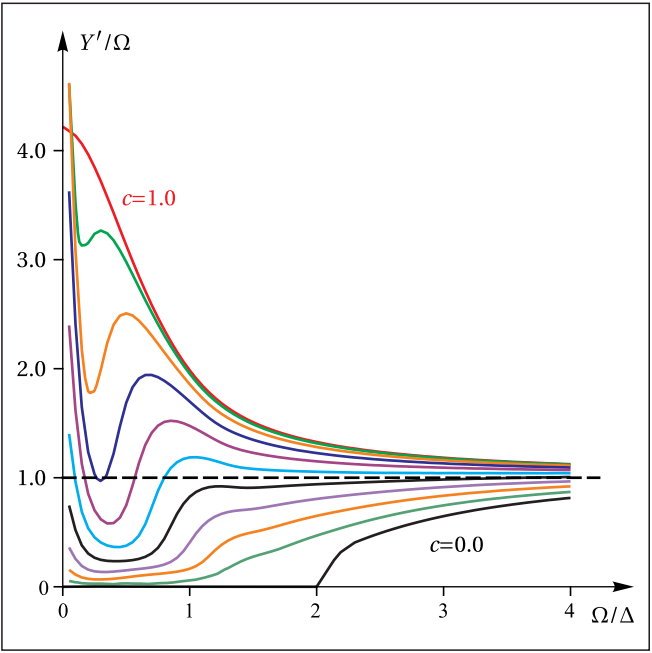}
\caption{(Color online.) The frequency dependence of the real part of admittance in case of S/N/S Josephson junction (scaled to its value in the normal state). Here, ${c = \cos \chi_{0}}$ and the values of~$c$ are incremented by~$0.1$ starting from~$0.0$ at bottom right and ending with~$1.0$ at top left. Note, that at small values of~$c$ the admittance is a monotonous function whereas for larger~$c$ a maximum emerges.}
\label{fig:Y}
\end{figure}

This means that the real part of admittance reaches a maximum at ${\Omega = 0}$ and the magnitude of this maximum is determined by the damping in the spectrum~$\gamma_{\text{N}}$ and temperature~$T$. At temperatures~$T$ below the subgap~$\Delta_{\text{sg}}$ the contribution of the anomalous current is exponentially small. In Fig.~\ref{fig:Y} we depict the frequency dependence of the real part of admittance ${Y_{\Omega}^{\prime} = \mathrm{Re} \{I_{\Omega} / V_{\Omega}\}}$ at different values~$\chi_{0}$.

\section{Triplet Coupling}
\label{sec:triplet_coupling}

In this Section, we consider a system with ``magnetic'' superconductors S$_{m}$ as reservoirs. As we noted above, these superconductors may be bulk superconductors~S covered by a thin layer of a weak ferromagnet~F with an exchange field~$\mathbf{h}$ oriented in the $x$~or $y$~direction, see Fig.~\ref{fig:Scheme}~(b). The triplet Cooper pairs that penetrate into the ferromagnetic layers have the total spin lying in the $y$\nobreakdash-$z$ or in $x$\nobreakdash-$z$~planes. The chosen form of the matrix element of tunneling through the filters~Fl [see~Eq.~(\ref{2})], $\mathrm{\check{\Gamma}}$, implies that only fully polarized in the $z$~direction triplet Cooper pairs can pass through the filters. Triplet Cooper pairs penetrating into the N~film from the left and from the right differ not only by their spin polarization, but also by their chiralities. The chiralities are not equal if the $\mathbf{h}$~vectors in the left and right reservoirs have different orientations, $\mathbf{h}||\mathbf{e}_{x}$ or $\mathbf{h}||\mathbf{e}_{y}$. Again, we consider first a stationary case.

\subsection{Stationary case}

We take the element $\{11\}$ of Eq.~(\ref{3}), that is, the equation for the retarded Green's function~$\hat{g}^{R}$. First, we write this equation for a stationary case in the Matsubara representation, i.e., we set ${\epsilon = i \omega}$, and come to
\begin{equation}
\omega \big[ \mathrm{\hat{X}}_{30} \,, \hat{g}_{\omega} \big] = E_{0} \big[ \hat{g}_{\omega} \,, \hat{G} \big] \,, \label{st1}
\end{equation}
where ${\hat{G} = \hat{G}_{||} + \hat{G}_{\perp}}$ with ${\mathrm{\hat{G}}_{||} = G_{\omega h} \big( \mathrm{\hat{X}}_{30} + s \mathrm{\hat{X}}_{03} \big)}$, ${s = (s_{\text{R}} + s_{\text{L}}) / 2}$, and the condensate Green's function
\begin{equation}
\hat{G}_{\perp} = 2^{-1} F_{\omega h}\big[ \big( \mathrm{\hat{X}}_{\perp \text{R}} + \mathrm{\hat{X}}_{\perp \text{L}} \big) \cos \chi_{0} + i \mathrm{\hat{X}}_{30} \big( \mathrm{\hat{X}}_{\perp \text{R}} - \mathrm{\hat{X}}_{\perp \text{L}} \big) \sin \chi_{0} \big] \,. \label{st1a}
\end{equation}
The functions~$G_{\omega h}$ and~$F_{\omega h}$ are defined as ${G_{\omega h} = G_{\omega +}}$ and~${F_{\omega h} = F_{\omega -}}$ in Eqs.~(\ref{5}) and~(\ref{5'}).

In the case of equal chiralities we have ${\mathrm{\hat{X}}_{\perp \text{R}} = \mathrm{\hat{X}}_{\perp \text{L}}}$. If the chiralities at the left and right~S$_{m}$ are different, we have ${\mathrm{\hat{X}}_{\perp \text{R}} \neq \mathrm{\hat{X}}_{\perp \text{L}}}$, where the form of the matrices $\mathrm{\hat{X}}_{\perp \text{R},\text{L}}$ is given by Eq.~(\ref{st3}).

Fisrt we consider the case of identical chiralities, i.e., ${\mathrm{\hat{X}}_{\perp \text{R}} = \mathrm{\hat{X}}_{\perp \text{L}}}$, and polarizations (${s_{\text{R}} = s_{\text{L}}}$).

\begin{enumerate}
\item[a)] Identical chiralities. In this case, the second term in Eq.~(\ref{st1a}) vanishes. If the polarizations on the left and on the right are opposite, the Josephson current is zero.\cite{Moor_Volkov_Efetov_2016_a} A solution of Eq.~(\ref{st1}) is searched in the form
\begin{equation}
\hat{g}_{\omega} = a_{30} \mathrm{\hat{X}}_{30} + a_{03} \mathrm{\hat{X}}_{03} + b_{\perp} \mathrm{\hat{X}}_{\perp \text{R}} \,. \label{st4}
\end{equation}

From Eq.~(\ref{st1}) we find
\begin{equation}
(\omega + 2 E_{0} G_{\omega h}) b_{\perp} = E_{0} F_{\omega h}(a_{30} + s a_{03}) \cos \chi_{0} \,. \label{st5}
\end{equation}%
The normalization condition ${\hat{g}_{\omega} \cdot \hat{g}_{\omega} = \hat{1}}$ yields
\begin{align}
a_{30}^{2} + a_{03}^{2} + b_{\perp}^{2} &= 1 \,, \\
2 a_{30} a_{03} + s b_{\perp} &=0 \,.
\label{st6}
\end{align}

\begin{figure}[tbp]
\includegraphics[width=1.0\columnwidth]{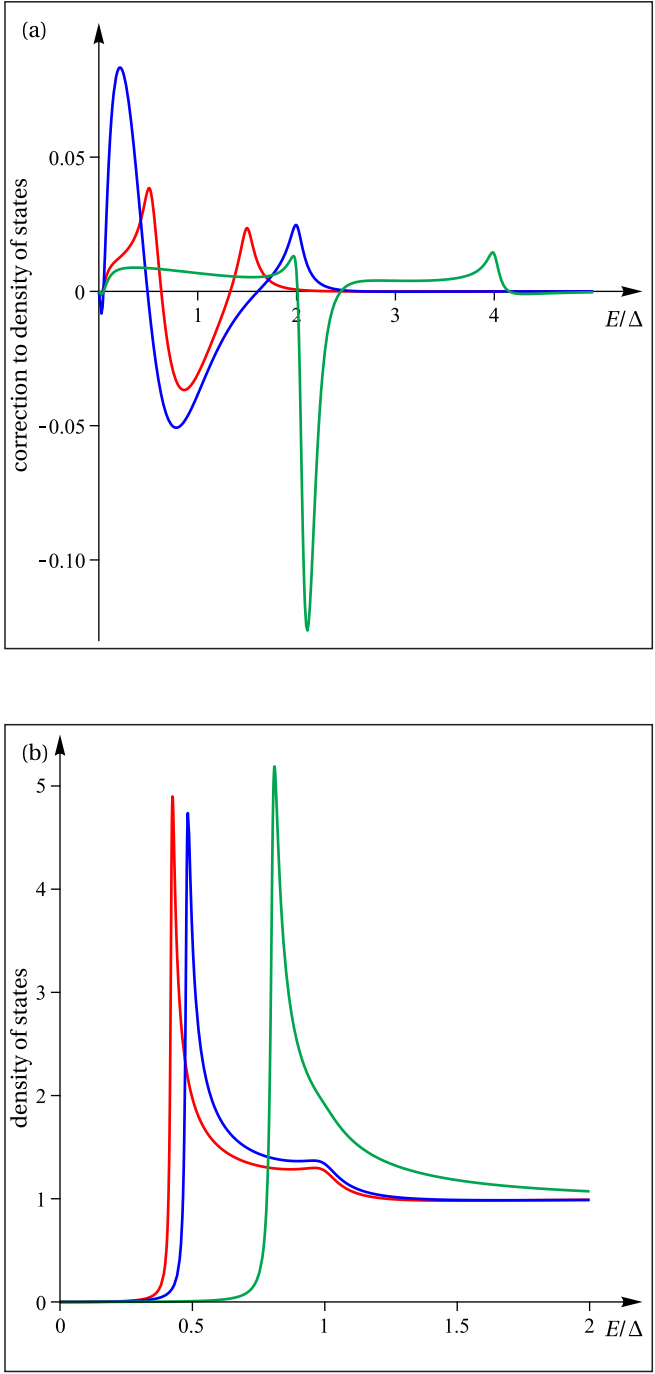}
\caption{(Color online.) The energy dependence of the DOS ${\nu(\epsilon) = \mathrm{Re} ( g_{\epsilon} )}$ for (a)~S$_{\text{m}}$/N/S$_{\text{m}}$, and (b)~S/N/S Josephson junctions. One can see that the correction to the DOS of the normal state is small. We set the value of cosine to ${c = 0.8}$ and, in the case~(a) of an S$_{\text{m}}$/N/S$_{\text{m}}$~contact, the curves correspond to ${h = 0.5 \Delta}$~(red), ${h = 1.0 \Delta}$~(blue), and ${h = 3.0 \Delta}$~(green); in case~(b) of an S$_{\text{m}}$/N/S$_{\text{m}}$~contact we vary the parameter~$E_0$ and the curves correspond to ${E_0 = 0.8 \Delta}$~(red), ${E_0 = 1.0 \Delta}$~(blue), and ${E_0 = 5.0 \Delta}$~(green).}
\label{fig:DOS STS}
\end{figure}

We find from Eqs.~(\ref{st5})\nobreakdash--(\ref{st6})
\begin{equation}
b_{\perp} = \frac{R}{\sqrt{1 + 2 R^{2}}} \,, \label{st7}
\end{equation}
and
\begin{align}
a_{30} &= 1 + s a_{03} \,, \\
s a_{03} &= -\frac{R^{2}}{A[1 + A]} \,, \label{st7'}
\end{align}
where ${A = \sqrt{1 + 2 R^{2}}}$ and ${R = (\omega + 2 E_{0} G_{\omega h})^{-1} (E_{0} F_{\omega h}) \cos \chi_{0}}$.

The parameter~$s a_{03}$ determines a correction to the DOS of the N~region $\delta \nu(\epsilon)$ [${\delta \nu(\epsilon) = s a_{03}(\omega)}$ at ${\omega = -i \epsilon}$] due to proximity effect. In Fig.~\ref{fig:DOS STS} we plot the DOS for ${E_{0}/\Delta = 1}$, ${\cos \chi_{0} = 0.8}$ and different~$h$. We see that the correction to the DOS of the normal metal [${\nu_{\text{N}}(\epsilon) = 1}$] is small and, consequently, there is no gap in the spectrum.\cite{BVErmp,GolubovPRL07} For comparison, we plot also the DOS of the considered singlet JJ where there is a gap $\Delta_{\text{sg}}$ in the excitation spectrum.

\begin{figure}[tbp]
\includegraphics[width=1.0\columnwidth]{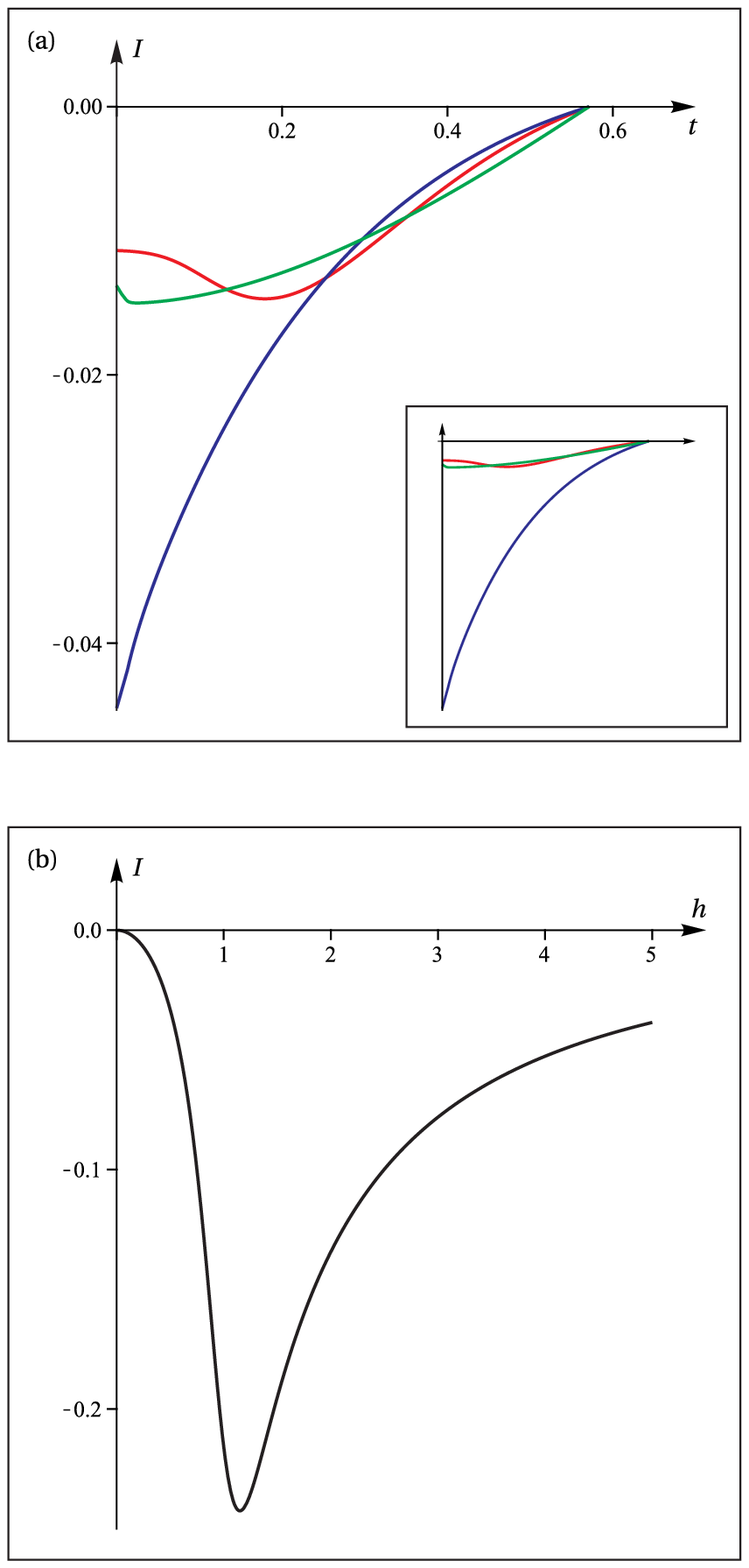}
\caption{(Color online.) The Josephson critical current~$I_{\text{T}, a}$ for the JJs with triplet coupling as a function of temperature~${t = T/\Delta}$~(a) and exchange field~$h$~(b). The temperature dependence of~$I_{\text{T}, a}$ is not monotonous at some values of the exchange field~$h$.\cite{Eschrig08,Grein_et_al_2009} In plotting the temperature dependence in~(a) we set ${E_0 = 0.5 \Delta}$, ${c = 0.5}$, and the curves correspond to ${h = 0.5 \Delta}$~(red), ${h = 1.0 \Delta}$~(blue), and ${h = 3.0 \Delta}$~(green). Note that the blue curve corresponding to ${h = 1.0 \Delta}$ is scaled by the factor of~$0.3$ and the unscaled curve is provided in the inset. Plotting the $h$~dependence we set ${E_0 = 0.5 \Delta}$, ${c = 0.5}$, and ${T = 0.1 \Delta}$.}
\label{fig:I_c}
\end{figure}

Expression Eq.~(\ref{st7}) allows us to find the Josephson current~$I_{\text{J}}$,
\begin{equation}
I_{\text{J}} = i \frac{\pi T}{e R_{\Box}} \sum_{\omega \geq 0} \big\{ \big[ \hat{g}_{\omega} \,, \hat{G}_{\omega \text{R}} \big] \big\}_{30} \,, \label{st8}
\end{equation}
where
\begin{equation}
\hat{G}_{\omega \text{R}} = G_{\omega h} \big( \mathrm{\hat{X}}_{30} + s \mathrm{\hat{X}}_{03} \big) + F_{\omega h} \big( \cos \chi_{0} + i \mathrm{\hat{X}}_{30} \sin \chi_{0} \big) \mathrm{\hat{X}}_{\perp \text{R}} \,. \label{st9}
\end{equation}

We find
\begin{align}
I_{\text{J}, a} &= I_{\text{T}, a} \sin (2 \chi_{0}) \,, \label{st10} \\
I_{\text{T}, a} &= \frac{\pi T}{2 e R_{\Box}} E_{0} \sum_{n \geq 0}^{\infty} \frac{F_{\omega h}^{2}}{\sqrt{\big(\omega + 2 E_{0} G_{\omega h}\big)^{2} + 2 \big(E_{0} F_{\omega h} \cos \chi_{0}\big)^{2}}} \,. \label{st10'}
\end{align}

When the total spins of triplet Cooper pairs stemming from the right (left) reservoirs are oriented in the same direction, Eq.~(\ref{st10'}) determines the critical currents of a $\pi$\nobreakdash-Josephson junction since~$F_{\omega h}$ is an imaginary quantity so that ${F_{\omega h}^{2} < 0}$. In Fig.~\ref{fig:I_c} we plot the temperature and~$h$ dependence of the critical current~$I_{T,a}$. We see that the temperature dependence of the critical current~$I_{\text{c}}(T)$ is not monotonous for some values of the exchange field~$h$.\cite{Eschrig08,Grein_et_al_2009}. As a function of~$h$, the absolute value of the critical current~$|I_{T}|$ increases with~$h$ from zero and reaches a maximum at a finite value of~$h_{\text{m}}$~[Fig.~\ref{fig:I_c}~(b)].

\item[b)] Different chiralities. In the case of different chiralities but equal polarizations the solution is found in a similar way. We look for the matrix~$\hat{g}_{\omega }$ in the form
\begin{equation}
\hat{g}_{\omega} = a_{30} \mathrm{\hat{X}}_{30} + a_{03} \mathrm{\hat{X}}_{03} + B_{\perp \text{R}} \mathrm{\hat{X}}_{\perp \text{R}} + B_{\perp \text{L}} \mathrm{\hat{X}}_{\perp \text{L}} \,.
\label{st11}
\end{equation}
We find ${B_{\perp \text{R}} = B_{\perp \text{L}} \equiv B_{\perp}}$ with
\begin{equation}
B_{\perp} = \frac{R \big(\cos \chi_{0} - s \sin \chi_{0}\big)}{\sqrt{1 + 4 R^{2}\big(\cos \chi_{0} - s \sin \chi_{0}\big)^{2}}} \,. \label{st11'}
\end{equation}

The current is determined by Eq.~(\ref{st8}) and we find
\begin{widetext}
\begin{align}
I_{\text{J}, b} &= I_{\text{T} b} \cos (2 \chi_{0}) \,, \label{st12} \\
I_{\text{T}, b} &= s \frac{\pi T}{2 e R_{\Box}} \sum_{n \geq 0}^{\infty} \frac{F_{\omega h}^{2}}{\sqrt{\big(\omega + 2 E_{0} G_{\omega h}\big)^{2} + \big(E_{0} F_{\omega h}\big)^{2}\big(\cos 2 \chi_{0} - s \sin 2 \chi_{0}\big)}} \,. \label{st12'}
\end{align}
\end{widetext}

In this case, the Josephson current appears even in the absence of a phase difference (the so-called anomalous current) and on the polarization of triplet Cooper pairs.\cite{Braude07,Grein_et_al_2009,Brydon09,Chan10,Margaris10,Linder14,Moor_Volkov_Efetov_2015_c,Moor_Volkov_Efetov_2016_a,Mironov_Buzdin_2015,Bergeret17} One can easily prove that the Josephson current is zero if the triplet components injected from the right and left S$_{\text{m}}$ have opposite spin orientations (${s_{\text{R}} = -s_{\text{L}} = \pm 1}$).\cite{Moor_Volkov_Efetov_2015_c,Moor_Volkov_Efetov_2016_a} It is interesting that at a given phase difference~$2 \chi_{0}$, the direction of the Josephson current depends on the spin polarization direction of triplet Cooper pairs, i.e., it is determined by the sign of~$s$. This situation is analogous to the case of superconductors with a special type of spin-orbit interaction where the direction of electron motion depends on the spin
polarization.\cite{Barash04,BuzdinPRL17}

As it was obtained in Ref.~\onlinecite{Moor_Volkov_Efetov_2015_c} (see also Ref.~\onlinecite{Bergeret17}), the quasiclassical approach leads to this correct result if the spin selection is realized through filters. If the filters are replaced by strong ferromagnets, triplet Cooper pairs described in the quasiclassical approximation penetrate into the N~region independently of the spin direction and the current~$I_{\text{T}}$ is not zero for both spin directions.
\end{enumerate}

Consider now a linear response of the system to an applied small ac voltage~$\delta V(t)$.

\subsection{Non-stationary case}

In the case of triplet coupling considered in this section we are interested in the low frequency range where only the anomalous Green's function~$\hat{g}^{\text{an}}$ are essential. We assume also that chiralities of the triplet components penetrating the N~region are identical, i.e., ${\mathrm{\hat{X}}_{\perp \text{R}} = \mathrm{\hat{X}}_{\perp \text{L}} = \mathrm{\hat{X}}_{11} - s \mathrm{\hat{X}}_{22}}$.

The anomalous Green's function~$\hat{g}^{\text{an}}$ obeys the linearized Eq.~(\ref{3a}),
\begin{equation}
\zeta_{+}^{R} \hat{g}_{\text{n}}^{R} \hat{g}^{\text{an}} - \hat{g}^{\text{an}} \hat{g}_{\text{n}}^{A} \zeta_{-}^{A} = i E_{0}\big[\hat{g}_{0}^{R} \hat{G}^{\text{an}} - \hat{G}^{\text{an}} \hat{g}_{0}^{A} - \hat{G}_{+}^{R} \hat{g}^{\text{an}} + \hat{g}^{\text{an}} \hat{G}_{-}^{A}\big] \,, \label{N1}
\end{equation}
where the matrices ${\hat{g}_{\text{n}}^{R(A)} = \pm \mathrm{\hat{X}}_{30}}$ are the matrix Green's functions in the N~region in the absence of the proximity effect, ${\zeta^{R(A)}(\bar{\epsilon}) = \bar{\epsilon} \pm i \gamma_{\text{N}}}$, that is, we take into account the damping~$\gamma_{\text{N}}$, ${\epsilon_{\pm} = \bar{\epsilon} \pm \Omega / 2}$, and $\Omega$ is the frequency of the phase variation ${\chi_{\Omega}(t) = \chi_{\Omega} \cos (\Omega t)}$. The matrices~$\hat{g}_{0}^{R(A)}$ describe the N(F)~wire in the stationary state. The obtained results for the N~region remain valid also for the F~region if the magnetization vector~$\mathbf{M}$ is oriented along the $z$~axis.

The method of finding a solution for a non-stationary equation, used in Section~\ref{sec:singlet_coupling}, is not applicable directly to the case of triplet coupling. We employ here a perturbation method regarding~$E_{0}$ as a small parameter and representing~$\hat{g}^{\text{an}}$ as a series in the powers of $E_{0}$, i.e., ${\hat{g}^{\text{an}} = \hat{g}_{1}^{\text{an}} + \hat{g}_{2}^{\text{an}} + \hat{g}_{3}^{\text{an}} + \ldots}$. The anomalous current~$I_{\Omega}^{\text{an}}$ is determined by the third-order Green's function~$\hat{g}_{3}^{\text{an}}$ [see Eqs.~(\ref{24A})\nobreakdash--(\ref{26A}) in Appendix~\ref{app:triplet}]. We calculate the ``spectral'' current~$j_{3}^{\text{an}}$ with the aid of~$\hat{g}_{3}^{\text{an}}$. It has the form
\begin{align}
j_{3}^{\text{an}} &= -\big\{\hat{G}_{\text{R} +}^{R} \cdot \hat{g}_{3 \perp}^{\text{an}} - \hat{g}_{3 \perp}^{\text{an}} \cdot \hat{G}_{\text{R} -}^{A} \big\}_{30} \notag \\
&= \frac{\chi_{\Omega}}{4} \frac{E_{0}^{3}\big(\cos \chi_{0} \sin \chi_{0}\big)^{2}}{\zeta_{0+}^{R} + \zeta_{0-}^{A}}\big(F_{\text{T} +}^{R} + F_{\text{T} -}^{A}\big)^{2} B(\epsilon) \label{N8a} \,.
\end{align}
where the functions~$F_{\text{T} \pm}^{R(A)}$ are defined in Eq.~(\ref{9Aa}) and the function~$B$ in Eq.~(\ref{26A}). The anomalous part of the current is
\begin{equation}
I_{\Omega}^{\text{an}} = \frac{1}{16 \pi e R_{\Box}} \int d \bar{\epsilon} \, n(\bar{\epsilon}) j_{3 \perp}^{\text{an}} \,. \label{N9}
\end{equation}

Consider limiting cases of large and small exchange field~$h$.
\begin{enumerate}
\item[a)] ${h \gg \Delta}$. In this case,
\begin{equation}
F_{\text{T}}^{R(A)} = \frac{\Delta}{2} \Big[\frac{1}{\zeta(\epsilon + h)} - \frac{1}{\zeta(\epsilon - h)}\Big]^{R(A)} \simeq \mp \frac{\Delta \epsilon}{h^{2}} \,. \label{N10}
\end{equation}

Therefore, we have $F_{\text{T} +}^{R} + F_{\text{T} -}^{A} \simeq - \Delta \Omega / h^{2}$. Calculating the integral in Eq.~(\ref{N9}) at ${\{T, E_{0}\} \ll \Delta}$, we find for the admittance
\begin{equation}
Y_{\perp}^{\text{an}}(\Omega) R_{\Box} = -\frac{E_{0} \sin^{2} 2 \chi_{0}}{32 \hbar (- i \Omega + \gamma )} \Big(\frac{E_{0}}{\Delta}\Big)^{2} \Big(\frac{\hbar \Omega}{\Delta}\Big)^{2} \Big(\frac{\Delta}{h}\Big)^{8} \,. \label{N11}
\end{equation}
We see that there is no singularity in the admittance at small~$\Omega$. Since the condensate functions ~$F_{\text{T}}^{R/(A)}$  are small in this case, the anomalous contribution gives a small contribution to the admittance in the normal state~$(1 / 2 R_{\Box})$.

\item[b)] ${h \ll \Delta}$. In this case, the admittance is
\begin{equation}
Y(\Omega) R_{\Box} = -\frac{2 E_{0} \sin^{2} 2 \chi_{0}}{\hbar (- i \Omega + \gamma)} \Big( \frac{E_{0}}{\Delta} \Big)^{2} \Big( \frac{T}{\Delta} \Big)^{2} \Big(\frac{h}{\Delta} \Big)^{4} c_{1} \,, \label{N12}
\end{equation}
where ${c_{1} = \int_{0}^{\infty} dx \, (x / \cosh x)^{2} \approx 0.82}$.

A singularity exists in this case, but the amplitude of the maximum of~$Y^{\prime}(\Omega)$ at ${\Omega \to 0}$ is rather small as the parameter~$(h/\Delta)$ and~$(E_{0} / \Delta)$ are small.
\end{enumerate}

\section{Conclusions}
\label{sec:conclusions}

In this paper, we have analyzed static and dynamic properties of the S/N/S and S$_{\text{m}}$/N(F)/S$_{\text{m}}$ Josephson weak links. In particular, the dc~Josephson current~$I_{\text{J}}$ was calculated as a function of temperature and of the parameter ${E_{0} / \Delta(0) = [E_{\text{Th}} / \Delta(0)] [R_{L}/R_{\Box}]}$, where ${E_{\text{Th}} = D / (2L)^{2}}$ is the Thouless energy and $R_{L} / R_{\Box}$ is the ratio of the resistance~$R_{L}$ of the N~film (or wire) with dimension~$L$ and of the interface resistance (per unit area)~$R_{\Box}$. The factor~$R_{L}/R_{\Box}$ is assumed to be small, whereas~$E_{\text{Th}} / \Delta(0)$---large, so that their product can be arbitrary. The coupling in the S/N/S and S$_{\text{m}}$/N/S$_{\text{m}}$ junctions is realized through the singlet and fully polarized triplet Cooper pairs, respectively. In this case, the term in the Hamiltonian,~$\mathbf{h}_{\text{str}} \hat{X}_{33}$, which describes the action of an exchange field~$\mathbf{h}_{\text{str}}$ on the spins of electrons commute with matrices in Eq.~(\ref{st3}) related to the triplet Cooper pairs. Therefore, for these Cooper pairs (long-range component) it doesn't matter whether the bridge consists of a normal metal or of a ferromagnet.

The superconducting ``magnetic'' reservoirs S$_{\text{m}}$ consist of conventional BCS superconductors covered by a thin F~layer with an exchange field~$\mathbf{h}$. Spin-filters at S$_{\text{m}}$/N interfaces let pass only Cooper pairs with the spin oriented along the $z$~axis. The Josephson current~$I_{\text{J}}$ is zero if Cooper pairs penetrating the N~(or~F)~region from the left and right reservoirs have opposite polarizations. The form of the dependence of~$I_{\text{J}}$ on the phase difference~$2 \chi_{0}$ is determined not only by relative polarization of triplet Cooper pairs, but also by the so called chiralities, that is, by the mutual directions of the vectors~$\mathbf{h}_{\text{R/L}}$ in the right and left S$_{\text{m}}$~reservoirs. In the case of identical chiralities (${\mathbf{h}_{\text{R}} \mathbf{h}_{\text{L}} / h_{\text{R}} h_{\text{L}} = 1}$), the Josephson current has the standard form ${I_{\text{J}} = I_{\text{T}, a} \sin (2 \chi_{0})}$, whereas in the case of different chiralities (${\mathbf{h}_{\text{R}} \mathbf{h}_{\text{L}} = 0}$), the phase difference dependence of~${I_{\text{J}} = I_{\text{T}, b} \cos (2 \chi_{0})}$, is rather unusual. This means that in the second case the Josephson current flows in the absence of the phase difference.\cite{Braude07,Grein_et_al_2009,Chan10,Margaris10,Linder14,Moor_Volkov_Efetov_2015_c,Moor_Volkov_Efetov_2016_a,Bergeret17}

Interestingly, the current~$I_{\text{T}, b}$ changes sign by inversion of the polarization direction, i.e., the direction of the Josephson current at a given phase difference~$(2 \chi_{0})$ is determined by polarization of the triplet Cooper pairs. This means that one can change the direction of the Josephson current by reversing the polarization of the spin filters. This phenomenon, to some extent, is similar to the spin-orbit interaction when the direction of electron motion depends on its spin polarization.\cite{Barash04,BuzdinPRL17} On the other hand, by changing the direction of the $\mathbf{h}_{\text{R},\text{L}}$~vectors at ${\chi_{0} = 0}$, one can switch the current~$I_{\text{J}}$ on or off since the chirality is changed. Therefore, one can control the charge current by switching the polarization of the weak ferromagnet~$F_{\text{w}}$.

We have also calculated the admittance~$Y(\Omega)$ for both singlet and triplet JJs in the presence of the dc Josephson current. The dependence of~$Y(\Omega)$ for the singlet JJs has been found for all frequencies. At low temperatures~$T$, the real part of admittance~$Y^{\prime}(\Omega)$ starts to increase at ${\Omega \geq \Delta_{\text{sg}}}$, but has a sharp peak at low frequencies if~$T$ is not too low. The anomalous behavior of~$Y^{\prime}(\Omega)$ is related to the presence of a gap and to a contribution of quasiparticles in the energy interval ${\Delta_{\text{sg}} < \epsilon < \Delta}$. Such anomalous behavior of~$Y^{\prime}(\Omega)$ was shown to be absent in triplet JJs with fully polarized Cooper pairs because there is no gap in these JJs. Thus, measuring the real part of the admittance~$Y^{\prime}(\Omega)$ at low frequencies, one can extract useful information about the odd-frequency triplet component in S$_{\text{m}}$/N/S$_{\text{m}}$ Josephson weak links.


\bigskip

\appendix

\section{Singlet Josephson junction}
\label{app:singlet}

The Green's functions in the right (left) reservoirs,~$\hat{G}_{\text{R},\text{L}}$, in a stationary case read
\begin{equation}
\hat{G}_{\text{R},\text{L}}^{R(A)} = \big[G_{0} \mathrm{\hat{X}}_{30} + i F_{0} \big(\cos \chi_{0} \pm i \mathrm{\hat{X}}_{30} \sin \chi_{0}\big) \big] \mathrm{\hat{X}}_{10}^{R(A)} \,, \label{1A}
\end{equation}
where~$2 \chi_{0}$ is the phase difference between superconducting reservoirs. The Green's function in Eq.~(\ref{B11}) is defined as follows
\begin{equation}
\hat{G}^{R(A)} = (1/2)\big(\hat{G}_{\text{R}} + \hat{G}_{\text{L}}\big)^{R(A)} \,. \label{2A}
\end{equation}
With the help of Eq.~(\ref{1A}) this equation can be written as
\begin{equation}
\hat{G}^{R(A)} = \big[G_{0} \mathrm{\hat{X}}_{30} + i F_{0} \cos \chi_{0} \mathrm{\hat{X}}_{10} \big]^{R(A)} \,. \label{3A}
\end{equation}

The variations of the Green's functions~$\delta \hat{G}_{\text{R},\text{L}}^{R(A)}$ due to ac phase~$\chi_{\Omega}$ are
\begin{widetext}
\begin{align}
\delta \hat{G}_{\text{R},\text{L}}^{R(A)} &= \delta \big\{\exp (i \chi_{\Omega} \mathrm{\hat{X}}_{30} / 2) \big[ G_{0} \mathrm{\hat{X}}_{30} + i F_{0} \big(\cos \chi_{0} \pm i \mathrm{\hat{X}}_{30} \sin \chi_{0}\big) \mathrm{\hat{X}}_{10} \big]^{R(A)} \exp (-i \chi_{\Omega} \mathrm{\hat{X}}_{30} / 2) \big\}  \label{3Aa}
\intertext{or}
\delta \hat{G}_{\text{R,}\text{L}}^{R(A)} &= \frac{i \chi_{\Omega}}{2} \big[ \big(G_{0-} - G_{0+}\big) \mathrm{\hat{1}} + i \big(F_{0+} + F_{0-}\big) \big(\mathrm{\hat{X}}_{30} \cos \chi_{0} \pm i \mathrm{\hat{1}} \sin \chi_{0}\big) \mathrm{\hat{X}}_{10} \big]^{R(A)} \,, \label{3Ab}
\end{align}
\end{widetext}
so that
\begin{equation}
\delta \hat{G}^{R(A)} = -\frac{i \chi_{\Omega}}{2} \big[ \big(F_{0+} + F_{0-}\big)^{R(A)} \sin \chi_{0} \big] \mathrm{\hat{X}}_{10} \,. \label{3Ac}
\end{equation}

The anomalous Green's function~$\hat{g}^{\text{an}}$ is determined by Eq.~(\ref{B16}), where the functions~$\hat{g}_{0}^{R(A)}$ can be easily obtained from Eq.~(\ref{B2}),
\begin{equation}
\hat{g}_{0}^{R(A)}(\epsilon) = \big[ \tilde{g}(\epsilon) \mathrm{\hat{X}}_{30} + i \tilde{f}(\epsilon) \mathrm{\hat{X}}_{10} \big]^{R(A)} \,, \label{4A}
\end{equation}
where ${\tilde{g} = \tilde{\epsilon} / \tilde{\zeta}(\tilde{\epsilon})}$, ${\tilde{f} = \tilde{\Delta}^{2}(\tilde{\epsilon}) / \tilde{\zeta}(\tilde{\epsilon})}$, ${\tilde{\zeta}(\tilde{\epsilon}) = \sqrt{\tilde{\epsilon}^{2} - \tilde{\Delta}^{2}(\tilde{\epsilon})}}$, ${\tilde{\epsilon} = \epsilon \big[1 + i E_{0} / \zeta(\epsilon) \big]}$, ${\tilde{\Delta}(\epsilon) = \big[ i E_{0} \Delta / \zeta(\epsilon) \big] \cos \chi_{0}}$ with ${\zeta(\epsilon) = \sqrt{\epsilon^{2} - \Delta^{2}}}$.
Anomalous Green's function in reservoirs $\hat{G}^{\text{an}}$ is defined according to
\begin{equation}
(T_{+} - T_{-}) \hat{G}^{\text{an}} = \delta \hat{G} - \delta \hat{G}^{R} T_{-} + T_{+} \delta\hat{G}^{A} \,, \label{5A}
\end{equation}
where ${T_{\pm} \equiv \tanh(\epsilon_{\pm} \beta)}$, the matrices~$\delta \hat{G}$ and~$\delta \hat{G}^{R(A)}$ are variations of the Keldysh and retarded (advanced) Green's functions in the presence of the ac phase variation ${\chi_{\Omega} = (\chi_{\Omega})_{\text{R}} = - ( \chi_{\Omega})_{\text{L}}}$. We obtain for~${\hat{G}_{\text{R},\text{L}}^{\text{an}}(\epsilon, \epsilon^{\prime}) = \hat{G}_{\text{R},\text{L}}^{\text{an}} 2 \pi \delta (\epsilon - \epsilon^{\prime} - \Omega)}$ with
\begin{align}
\hat{G}_{\text{R},\text{L}}^{\text{an}} &= \pm \frac{i \chi_{\Omega}}{2} \big\{ \mathrm{\hat{X}}_{30} \cdot \big[ \big( \hat{G}_{-}^{R} - \hat{G}_{-}^{A} \big) T_{-} - \hat{G}_{-}^{R} T_{-} + T_{+} \hat{G}_{-}^{A} \big]  \label{6A} \\
&- \big[ \big( \hat{G}_{+}^{R} - \hat{G}_{+}^{A} \big) T_{+} - \hat{G}_{+}^{R} T_{-} + T_{+} \hat{G}_{+}^{A} \big] \cdot \mathrm{\hat{X}}_{30} \big\} \,. \notag
\end{align}
Here, the matrices~${\hat{G}_{\pm}^{R(A)} \equiv \hat{G}_{0}^{R(A)}(\epsilon_{\pm})}$ are given in Eqs.~(\ref{3c}) and~(\ref{2A}). For the matrix ${\hat{G}^{\text{an}} = \big( \hat{G}_{\text{R}}^{\text{an}} + \hat{G}_{\text{L}}^{\text{an}} \big) / 2}$ we get from Eq.~(\ref{6A})
\begin{equation}
\hat{G}^{\text{an}} = - \frac{i \chi_{\Omega}}{2} \big( F_{+}^{R} + F_{-}^{A} \big) \mathrm{\hat{X}}_{10} \sin \chi_{0} \,. \label{7A}
\end{equation}
The functions ${F_{+}^{R} = F_{0}^{R}(\epsilon_{\pm})}$ and ${F_{0}^{R}(\epsilon)}$ are defined in Eq.~(\ref{3c}). Using Eqs.~(\ref{4A}) and~(\ref{7A}), we obtain for the anomalous Green's function~$\hat{g}^{\text{an}}$
\begin{equation}
\hat{g}^{\text{an}} = \frac{\chi_{\Omega} E_{0}}{2 \big( \tilde{\zeta}_{+}^{R} + \tilde{\zeta}_{-}^{A} \big)} \big( F_{+}^{R} + F_{-}^{A} \big) \big[1 + \tilde{g}_{+} \tilde{g}_{-} + \tilde{f}_{+} \tilde{f}_{-} \big] \mathrm{\hat{X}}_{10} \sin \chi_{0} \,. \label{8A}
\end{equation}

Using Eqs.~(\ref{1A}),~(\ref{4A}), and~(\ref{8A}), we obtain the expression Eq.~(\ref{B19}) for the current with~$j_{1}^{\text{an}}$ and~$j_{2}^{\text{an}}$ defined in Eqs.~(\ref{B20}) and~(\ref{B20a}).

\section{Triplet Josephson junction. Perturabative approach}
\label{app:triplet}

\subsection{Stationary case}

We consider the case of small energy~$E_{0}$ (${E_{0} \ll \Delta}$) and obtain corrections~$\delta \hat{g}_{0}^{R(A)}$ to the stationary Green's functions ${\hat{g}_{0}^{R(A)} \equiv \pm \mathrm{\hat{X}}_{30}}$ as well as the anomalous function~$\hat{g}^{\text{an}}$ in a non-stationary case.

Consider first a stationary case and represent the matrix~$\hat{g}_{\text{st}}^{R(A)}$ in the form of expansion in powers of~$E_{0}$, i.e., ${\hat{g}_{\text{st}}^{R(A)} = [\hat{g}_{0} + \delta_{1} \hat{g} + \delta _{2} \hat{g} + \ldots]^{R(A)}}$.

Equation~(\ref{st1}) can be written for~$\hat{g}_{\text{st}}^{R(A)}$ in the form [for brevity we drop the indices~$R(A)$]
\begin{equation}
\epsilon \big[ \mathrm{\hat{X}}_{30} \,, \hat{g}_{\text{st}} \big] = i E_{0} \big[ \hat{g}_{\text{st}} \,, \hat{G}_{0} \big] \,, \label{9A}
\end{equation}
where the matrix~$\hat{G}_{0}$ is
\begin{equation}
\hat{G}_{0} = G_{\text{T}} \mathrm{\hat{X}}_{30} + i F_{\text{T}} \big(\cos \chi_{0} + i \sin \chi_{0} \mathrm{\hat{X}}_{30}\big) \mathrm{\hat{X}}_{\perp} \label{9Aa}
\end{equation}
with ${G_{\text{T}} = 2^{-1} \big[ G_{0}(\epsilon + h) + G_{0}(\epsilon - h) \big]}$, ${F_{\text{T}} = 2^{-1} \big[F_{0}(\epsilon + h) - F_{0}(\epsilon - h)\big]}$, and the functions~$G_{0}^{R(A)}$ and~$F_{0}^{R(A)}$ are defined in Eq.~(\ref{3c}). The first-order correction obeys the equation
\begin{equation}
\epsilon \big[ \mathrm{\hat{X}}_{30} \,, \delta_{1} \hat{g}\big] = i E_{0} \big[\hat{g}_{0} \,, \hat{G}_{0}\big] \,, \label{10A}
\end{equation}
and the solution is
\begin{equation}
\delta_{1} \hat{g} = - E_{0} \cos \chi_{0} \Big(\frac{F_{\text{T}}}{\zeta_{0}} \Big) \mathrm{\hat{X}}_{\perp} \label{11A}
\end{equation}
with ${\zeta_{0}^{R(A)} = \pm (\epsilon \pm i \gamma)}$. The second-order correction~$\delta_{2} \hat{g}$ satisfies the equation
\begin{equation}
\epsilon \big[ \mathrm{\hat{X}}_{30} \,, \delta_{2} \hat{g}\big] = i E_{0} \big[ \delta_{1} \hat{g} \,, \hat{G}_{0} \big] \,. \label{12A}
\end{equation}
The normalization condition yields
\begin{equation}
\hat{g}_{0} \cdot \delta_{2} \hat{g} + \delta_{2} \hat{g} \cdot \hat{g}_{0} + (\delta_{1} \hat{g})^{2} = 0 \,. \label{13A}
\end{equation}

The solution satisfying Eqs.~(\ref{12A}) and~(\ref{13A}) is represented in the form
\begin{equation}
\delta_{2} \hat{g} = a \mathrm{\hat{X}}_{30} + b\mathrm{\hat{X}}_{\perp} \label{14A}
\end{equation}
with
\begin{align}
a &= - \frac{(E_{0} \cos \chi _{0})^{2}}{2} \Big( \frac{F}{\zeta_{0} \zeta_{\Delta}} \Big)^{2} \,, \label{15A} \\
b &= i E_{0}^{2} \Big( \frac{F}{\zeta_{0}} \Big)^{2} \cos \chi_{0} \,,
\end{align}
where ${\zeta_{\Delta} = \sqrt{\epsilon^{2} - \Delta^{2}}}$.

\subsection{Non-stationary case}

Consider now a non-stationary case. In order to find the anomalous Green's function~$\hat{g}^{\text{an}}$, we use Eqs.~(\ref{11A})\nobreakdash--(\ref{14A}). This function obeys Eq.~(\ref{N1}), where the matrix~$\hat{G}^{\text{an}}$ looks similarly to that provided in Eq.~(\ref{6A}),
\begin{equation}
\hat{G}^{\text{an}} = - \frac{i \chi_{\Omega}}{2} \big[ F_{+}^{R} + F_{-}^{A} \big] \mathrm{\hat{X}}_{\perp} \sin \chi_{0} \,. \label{6A'}
\end{equation}

The first-order correction $\hat{g}_{1}^{\text{an}}$ obeys the equation
\begin{equation}
\epsilon_{+} \mathrm{\hat{X}}_{30} \cdot \hat{g}_{1}^{\text{an}} - \hat{g}_{1}^{\text{an}} \cdot \mathrm{\hat{X}}_{30} \epsilon_{-} = i E_{0} \big[ \hat{g}_{0 +} \cdot \hat{G}^{\text{an}} - \hat{G}^{\text{an}} \cdot \hat{g}_{0 -}\big] \,, \label{17A}
\end{equation}
where ${\hat{g}_{0 +} \equiv \hat{g}_{0 +}^{R}}$ and ${\hat{g}_{0 -} \equiv \hat{g}_{0 -}^{A}}$. Taking into account that ${\hat{g}_{0 +} = - \hat{g}_{0 -} = \mathrm{\hat{X}}_{30}}$ and Eq.~(\ref{7A}), we obtain that
\begin{equation}
\hat{g}_{1}^{\text{an}} = 0 \,. \label{18A}
\end{equation}

The second-order correction~$\hat{g}_{2}^{\text{an}}$ satisfies the equation
\begin{equation}
\zeta_{+} \hat{g}_{0 +} \cdot \hat{g}_{2}^{\text{an}} - \hat{g}_{2}^{\text{an}} \cdot \hat{g}_{0 -} \zeta_{-} = i E_{0} \big[ \delta_{1} \hat{g}_{+} \cdot \hat{G}^{\text{an}} - \hat{G}^{\text{an}} \cdot \delta_{1} \hat{g}_{-} \big] \,. \label{19A}
\end{equation}
We use the normalization condition
\begin{equation}
\hat{g}_{0 +} \cdot \hat{g}_{2}^{\text{an}} + \hat{g}_{2}^{\text{an}} \cdot \hat{g}_{0 -} = 0 \,, \label{19A'}
\end{equation}
so that the solution of Eq.~(\ref{19A}) is
\begin{equation}
\hat{g}_{2}^{\text{an}} = g_{2}^{\text{an}} \mathrm{\hat{X}}_{30} \,, \label{20A}
\end{equation}
where~$g_{2}^{\text{an}}$ is
\begin{equation}
g_{2}^{\text{an}} = \frac{i E_{0}}{\zeta_{+} + \zeta_{-}} G^{\text{an}} \big( \delta_{1} g_{+} - \delta_{1} g_{-} \big) \,. \label{21A}
\end{equation}

We need to find the third-order correction~$\hat{g}_{3}^{\text{an}}$ which obeys the equations
\begin{widetext}
\begin{align}
\zeta_{+} \hat{g}_{0 +} \cdot \hat{g}_{3}^{\text{an}} - \hat{g}_{3}^{\text{an}} \cdot \hat{g}_{0 -} \zeta_{-} &= i E_{0} \big[ \delta_{2} \hat{g}_{+} \cdot \hat{G}^{\text{an}} - \hat{G}^{\text{an}} \cdot \delta_{2} \hat{g}_{-} + \hat{g}_{2}^{\text{an}} \cdot \hat{G}_{0 -} - \hat{G}_{0 +} \cdot \hat{g}_{2}^{\text{an}} \big]  \label{22A} \\
\hat{g}_{0 +} \cdot \hat{g}_{3}^{\text{an}} + \hat{g}_{3}^{\text{an}} \cdot \hat{g}_{0 -} + \delta_{1} \hat{g}_{+} \cdot \hat{g}_{2}^{\text{an}} + \hat{g}_{2}^{\text{an}} \cdot \delta_{1} \hat{g}_{-} &= 0 \,. \label{23A}
\end{align}
\end{widetext}
The contribution to the current is given merely by the part~$\hat{g}_{3 \perp}^{\text{an}}$ of~$\hat{g}_{3}^{\text{an}}$ which is proportional to the matrix~$\mathrm{\hat{X}}_{\perp}$. We obtain for~$\hat{g}_{3 \perp}^{\text{an}}$
\begin{equation}
\hat{g}_{3 \perp}^{\text{an}} = g_{3 \perp}^{\text{an}} \mathrm{\hat{X}}_{\perp} \,, \label{24A}
\end{equation}
where~$g_{3 \perp}^{\text{an}}$ is given by
\begin{equation}
g_{3 \perp}^{\text{an}} = \frac{\chi_{\Omega}}{4} \frac{E_{0}^{3}(\cos \chi_{0})^{2} \sin \chi _{0}}{\zeta_{+} + \zeta _{-}} \big( F_{\text{T} +} + F_{\text{T} -} \big) B \,, \label{25A}
\end{equation}
and the function~$B$ is
\begin{equation}
B = \Big( \frac{F_{\text{T} +}}{\zeta_{+}} \Big)^{2} - 2 \Big( \frac{F_{\text{T} +} F_{\text{T} -}}{\zeta_{+} \zeta_{-}} \Big) - \Big( \frac{F_{\text{T} -}}{\zeta_{-}} \Big)^{2} \,. \label{26A}
\end{equation}

In obtaining Eqs.~(\ref{24A})\nobreakdash--(\ref{26A}) we used expressions for corrections~$\delta_{1} \hat{g}_{0}^{R(A)}$ and~$\delta_{2} \hat{g}_{0}^{R(A)}$ to the Green's functions~$\hat{g}_{\text{n}}^{R(A)}$ in the static case. Note that this formula is applicable both for the triplet and for the singlet JJ, S/N/S, because the only property we used is that the matrix
\begin{equation}
\mathrm{\hat{X}}_{\perp} = \begin{cases}
\mathrm{\hat{X}}_{11} - s \mathrm{\hat{X}}_{22} \,, & \text{triplet JJ} \,, \\
\mathrm{\hat{X}}_{10} \,, & \text{singlet JJ} \,,
\end{cases}
\label{27A}
\end{equation}
anticommutes with the matrix~$\mathrm{\hat{X}}_{30}$.

Using Eqs.~(\ref{24A})\nobreakdash--(\ref{26A}), we obtain Eqs.~(\ref{N8a})\nobreakdash--(\ref{N9}).

\section{Inverse Proximity Effect}
\label{app:inv_prox_eff}

Consider, for simplicity, a contact between a singlet superconductor~S and a normal metal~N shown in Fig.~\ref{fig:Scheme}~(a). The Green's functions in the superconductor~$\hat{G}_{\omega \text{S}}(z)$ in a stationary case obeys the equation
\begin{equation}
-D_{\text{S}} \partial_{z} (\hat{G}_{\omega \text{S}} \cdot \partial_{z} \hat{G}_{\omega \text{S}}) + \omega \big[ \mathrm{\hat{X}}_{30} \,, \hat{G}_{\omega \text{S}} \big] + \Delta \big[ \mathrm{\hat{X}}_{10} \,, \hat{G}_{\omega \text{S}} \big] = 0 \,. \label{1C}
\end{equation}
The boundary condition for $\hat{G}_{\omega \text{S}}$ is
\begin{equation}
\hat{G}_{\omega \text{S}} \cdot \partial_{z} \hat{G}_{\omega \text{S}} = \varkappa_{\text{S}} \big[\hat{G}_{\omega \text{S}} \,, \hat{g}_{\omega \text{N}} \big] \,, \label{2C}
\end{equation}
where ${\varkappa_{\text{S}} = (R_{\square} \sigma_{\text{S}})^{-1}}$. We integrate Eq.~(\ref{1C}) over the thickness~$d_{\text{S}}$ of the S~film assuming that the function~$\hat{G}_{\omega \text{S}}(z)$ is almost constant (later, we check this assumption) and taking into account the boundary condition Eq.~(\ref{2C}).

We obtain the equation
\begin{equation}
\big[ \hat{M}_{\text{S}} \,, \hat{G}_{\omega \text{S}} \big] = 0 \,, \label{3C}
\end{equation}
which looks similar to Eq.~(17). Here, ${\hat{M}_{\text{S}} = \tilde{\omega} \mathrm{\hat{X}}_{30} + \Delta \mathrm{\hat{X}}_{10}}$, ${\tilde{\omega} \simeq \omega + D_{\text{S}} \varkappa_{\text{S}} / d_{\text{S}}}$. The quantity ${\gamma \equiv D_{\text{S}} \varkappa_{\text{S}} / d_{\text{S}}}$ is a damping in the superconductor~S which is induced due to inverse proximity effect. This factor is small compared to~$\Delta$ if the condition ${D_{\text{S}} \varkappa_{\text{S}} / d_{\text{S}} \simeq \langle 1 - \mathcal{R} \rangle \sqrt{\Delta / \tau_{\text{S}}} \ll \Delta}$ is fulfilled, where ${\langle 1 - \mathcal{R} \rangle}$ is an average transmission coefficient of electron passage through the S/N~interface which is supposed to be small~(see Ref.~\onlinecite{Volkov199321}), and~$\mathcal{R}$ is the reflection coefficient. We assumed that ${d_{\text{S}} \simeq \xi_{\text{S}}}$, where~$\xi_{\text{S}}$ is the correlation length in the superconductor~S and~$\tau_{\text{S}}$ is the momentum relaxation time there. However, even if the condition above is not fulfilled, but the thickness~$d_{\text{S}}$ is larger than~$\xi_{\text{S}}$, the obtained results remain valid with a reduced value of~$\Delta$ at the S/N~interface [${\Delta(0) < \Delta(d_{\text{S}})}$].

\section{Formula for AC Current}
\label{app:ac_current_formulas}

We write Eq.~(\ref{6}) for the ac current~$\delta I(t)$ (for brevity we set ${\mathrm{\check{\Gamma}} = \check{1}}$ so that there is no difference between matrices~$\mathrm{\check{G}}$ and~$\check{G}$) in the form
\begin{align}
\delta I(t) &= (16 \kappa R_{\Box} e)^{-1} \, \int dt_{1} \, \big\{\big[\delta \check{g}(t,t_{1}) \,, \check{G}_{0}(t_{1}-t) \big]^{K} \label{4D} \\
&+ \big[\check{g}_{0}(t-t_{1}) \,, \delta \check{G}(t_{1},t) \big]^{K}\big\} \,. \notag
_{30}
\end{align}

Consider the first term in Eq.~(\ref{4D}) (the second term can be recast in the same way)
\begin{widetext}
\begin{align}
\delta I_{1}(t)(16 \kappa R_{\Box} e) &= \big\{ \big[ \delta \check{g}(t,t_{1}) \cdot \check{G}_{0}(t_{1}-t) - \check{G}_{0}(t-t_{1}) \cdot \delta \check{g}(t_{1},t) \big]^{K} \big\}_{30}  \label{5D} \\
&= \int dt_{1} \int d\epsilon \int \frac{d\epsilon_{1}}{2 \pi} \int \frac{d\epsilon_{2}}{2 \pi} \big\{ \big[ \delta \check{g}(\epsilon,\epsilon_{1}) \cdot \check{G}_{0}(\epsilon_{2}) \exp [-i \epsilon t + i \epsilon_{1} t_{1} - i \epsilon_{2} (t_{1} - t)] \notag \\
&- \check{G}_{0}(\epsilon) \cdot \delta \check{g}(\epsilon_{1},\epsilon_{2}) \exp [- i \epsilon (t-t_{1}) - i \epsilon_{1} t_{1} + i \epsilon_{2} t ] \big]^{K} \big\}_{30} \notag \\
&= \int d\epsilon \int \frac{d\epsilon_{1}}{2 \pi} \big\{ \big[ \delta \check{g}(\epsilon,\epsilon_{1}) \cdot \check{G}_{0}(\epsilon_{1}) - \check{G}_{0}(\epsilon) \cdot \delta \check{g}(\epsilon,\epsilon_{1}) \big]^{K} \big\}_{30} \exp [-i(\epsilon - \epsilon_{1}) t]  \,. \notag
\end{align}
\end{widetext}

Taking into account that ${\delta \check{g}(\epsilon,\epsilon_{1}) = 2 \pi \delta(\epsilon - \epsilon_{1} - \Omega) \delta \check{g}}$, we obtain
\begin{equation}
\delta I(t) = (16 \kappa R_{\Box} e)^{-1} \exp [-i \Omega t] I_{\Omega} \, \label{6D}
\end{equation}
where
\begin{widetext}
\begin{equation}
I_{\Omega} = (16 \kappa R_{\Box} e)^{-1} \int d\bar{\epsilon} \big\{ \big[ \delta \check{g} \cdot \check{G}_{0}(\epsilon_{-}) - \check{G}_{0}(\epsilon_{+}) \cdot \delta \check{g} \big]^{K} + \big[\check{g}(\epsilon_{+}) \cdot \delta \check{G} - \delta \check{G} \cdot \check{g}(\epsilon_{-}) \big]^{K} \big\}_{30} \,.  \label{7D}
\end{equation}
\end{widetext}

The current~$I_{\Omega}$ is related to the admittance according to the standard expression ${I_{\Omega} = Y(\Omega) V_{\Omega}}$.


%

\end{document}